\documentclass[conference]{IEEEtran}
\IEEEoverridecommandlockouts

\usepackage{cite}
\usepackage{algorithm}
\usepackage{amsmath,amssymb,amsfonts}
\usepackage{graphicx}
\usepackage{verbatim}
\usepackage{paralist}
\usepackage{textcomp}
\usepackage{xcolor}
\usepackage{colortbl}
\usepackage{array}
\usepackage{booktabs}
\usepackage{subfigure}
\usepackage{multirow}
\usepackage{amsthm}
\usepackage{pifont}
\newtheorem{example}{Example}
\usepackage{threeparttable}
\usepackage{algpseudocode}

\def\BibTeX{{\rm B\kern-.05em{\sc i\kern-.025em b}\kern-.08em
    T\kern-.1667em\lower.7ex\hbox{E}\kern-.125emX}}
\begin{document}

\title{Limit Order Book Event Stream Prediction with Diffusion Model}

\author{\IEEEauthorblockN{Zetao Zheng$^{\dagger}$ \quad Guoan Li$^{\dagger}$ \quad Deqiang Ouyang$^{\ddagger}$ \quad Decui Liang$^{\dagger}$ \quad Jie Shao$^{\dagger}$}
\IEEEauthorblockA{$^{\dagger}$\textit{University of Electronic Science and Technology of China, Chengdu, China}\\
$^{\ddagger}$\textit{Chongqing University, Chongqing, China}\\
\{ztzheng, dcliang, shaojie\}@uestc.edu.cn \quad
liguoan@std.uestc.edu.cn \quad deqiangouyang@cqu.edu.cn}}

\maketitle

\begin{abstract}
Limit order book (LOB) is a dynamic, event-driven system that
records real-time market demand and supply for a financial asset in
a stream flow. Event stream prediction in LOB refers to forecasting
both the timing and the type of events. The challenge lies in
modeling the time-event distribution to capture the interdependence
between time and event type, which has traditionally relied on
stochastic point processes. However, modeling complex market
dynamics using stochastic processes, e.g., Hawke stochastic process,
can be simplistic and struggle to capture the evolution of market
dynamics. In this study, we present LOBDIF (\textbf{LOB} event
stream prediction with \textbf{dif}fusion model), which offers a new
paradigm for event stream prediction within the LOB system. LOBDIF
learns the complex time-event distribution by leveraging a diffusion
model, which decomposes the time-event distribution into sequential
steps, with each step represented by a Gaussian distribution.
Additionally, we propose a denoising network and a skip-step
sampling strategy. The former facilitates effective learning of
time-event interdependence, while the latter accelerates the
sampling process during inference. By introducing a diffusion model,
our approach breaks away from traditional modeling paradigms,
offering novel insights and providing an effective and efficient
solution for learning the time-event distribution in order streams
within the LOB system. Extensive experiments using real-world data
from the limit order books of three widely traded assets confirm
that LOBDIF significantly outperforms current state-of-the-art
methods.
\end{abstract}


\section{Introduction}
\label{sec:introduction}

Most modern financial markets use the continuous double auction
(CDA) mechanism to determine asset prices. Buyers and sellers
interact through physical or digital venues by submitting bid orders
(i.e., buy orders) and ask orders (i.e., sell orders), which include
price and volume details, into a queuing system known as the limit
order book (LOB). A matching engine then pairs these orders into
transactions, and the resulting sequence of transaction prices
defines the asset's price at the micro level. Such order data can be
effectively represented as an irregularly sampled time series of
event stream generated by market participants, along with an
event-driven, continuously updated LOB that stores all unexecuted
orders in the market. The accumulation of these order streams drives
the evolution of the LOB system. Thus, understanding the dynamics of
order flow and stream patterns can enhance our ability to predict
LOB behavior, providing valuable insights into market depth and
potential price movements.

Modeling the event stream LOB presents significant challenges due to
its diverse event types, such as order submissions and cancellations
on both bid and ask sides, as well as its irregularity, with
asynchronous trading actions occurring at arbitrary time points by
market participants. A common approach to capturing the arrival of
events in LOB is through stochastic point processes, where intensity
functions control the frequency of event occurrences over a given
time interval. In a naive setting, intensity rates may be
independent of past events, as in a Poisson process
\cite{DBLP:journals/ior/ContST10}, whereas more sophisticated and
realistic models incorporate history-dependent intensity rates, such
as in Hawkes processes \cite{DBLP:journals/siamfm/AbergelJ15}.
Leveraging recent advancements in deep learning, neural point
processes have been introduced to model stochastic point processes
using neural networks \cite{DBLP:conf/nips/MeiE17,
DBLP:conf/icml/ZhangLKY20}. This integration enhances the predictive
accuracy of traditional stochastic models and has demonstrated its
state-of-the-art performance in LOB prediction
\cite{DBLP:journals/corr/abs-2207-09951, DBLP:conf/kdd/ShiC22,
DBLP:conf/pakdd/ChungLK24}.

However, current mainstream approaches for modeling LOB event stream
often rely on specific stochastic point processes
\cite{DBLP:conf/pakdd/ChungLK24, DBLP:conf/kdd/ShiC22}. These
methods predict event types based on intensity functions learned
from event histories. Despite their popularity, they face the
following challenges:
\begin{enumerate}
    \item \textbf{Simplistic distributional assumptions}: The evolution of the
LOB system is highly complex, and using a straightforward stochastic
point process (e.g., Poisson process or Hawkes process) often falls
short in capturing this complexity. For instance, the Hawkes process
models LOB dynamics through intensity functions and decaying
functions. However, this simplistic mechanism may struggle to
capture the complex dynamics and the continuously evolving nature of
the LOB system.
    \item \textbf{Entangled time-event joint distribution}: The time-event
joint distribution captures the interdependence between various
event occurrence time and types. Modeling these joint time-event
distributions for the event stream generally involves a
high-dimensional sample space, which makes it highly intractable in
practice.
    \item \textbf{Limited support for closed-form sampling}: Many stochastic
models, including the ones based on intensity functions, do not
easily support closed-form sampling, which refers to the ability to
directly sample from a distribution using a deterministic formula
without iterative procedures. This limitation hinders efficient and
scalable prediction, especially when dealing with large-scale LOB
data streams, where speed and accuracy are crucial.
\end{enumerate}

Addressing the above challenges requires a new paradigm for modeling
LOB event stream. To this end, this work proposes LOBDIF
(\textbf{LOB} event stream prediction with \textbf{dif}fusion
model), a diffusion-based model for event stream prediction in the
LOB system, which learns the complex time-event distribution by
decomposing it into a Markov chain of multiple steps. Each step
represents a small distributional change that can be accurately
modeled by a Gaussian distribution
\cite{DBLP:conf/icml/Sohl-DicksteinW15}. The target distribution is
learned through the aggregation of all steps, with the predicted
joint distribution from the previous step serving as the condition
for the subsequent step. This structure enables closed-form
sampling, as each step involves direct sampling from a Gaussian
distribution without requiring iterative approximations, making the
sampling process both efficient and tractable. By breaking down the
complex learning process into a sequence of simpler steps involving
Gaussian distributions, LOBDIF captures the evolution of the event
stream more accurately and makes the intractable time-event joint
distribution more manageable, effectively addressing the challenges
faced by previous methods.

However, introducing a diffusion model in LOB modeling presents two
significant technical challenges: (1) \textbf{Ineffectiveness in
modeling time-event relationships}. Capturing the relationship
between events at each step of the diffusion process is crucial, as
it directly impacts the effectiveness of the final time-event
distribution. To address this, a denoising network is designed to
capture relationships between time and events. Specifically, time
attention and event attention are learned simultaneously to
adaptively capture fine-grained interactions, characterizing the
underlying mechanisms of the joint distribution. (2) \textbf{Low
sampling efficiency}. In real-world LOB systems, events occur at
high frequency, often requiring rapid decision-making. However, the
diffusion model relies on an iterative denoising process to generate
samples, which is inherently slow and inefficient for LOB
applications. To address this, a skip-step sampling strategy is
introduced, which skips some denoising steps during the reverse
denoising process, effectively reducing the number of iterations
required. This strategy allows for faster sampling without
compromising prediction accuracy, significantly enhancing sampling
efficiency and making it more suitable for real-time LOB
applications.

This paper presents the first systematic attempt to utilize
diffusion model in LOB system event stream prediction. The main
contributions are three-fold:
\begin{compactitem}
\item We propose a new paradigm for modeling the event stream in
LOB to address the common challenges faced by previous approaches
that rely on specific stochastic point processes.
\item A denoising network and a skip-step sampling strategy are introduced. The former
facilitates learning the time-event joint distribution, while the
latter ensures efficient sampling during the denoising phase.
\item Experiments on three real-world market datasets
demonstrate our model's effectiveness and efficiency.
\end{compactitem}

\section{Related Work}
\label{sec:rw}

\subsection{Modeling of LOB Event Stream}

Traditional models for modeling LOB event stream dynamics can be
broadly categorized into \textit{stochastic models} and
\textit{equilibrium models}. Stochastic models are widely used to
capture the probabilistic nature of order stream, typically assuming
that events such as the arrival and cancellation of limit orders
follow specific probabilistic processes. Some studies model LOB
evolution as a higher-order Markov system, where events follow a
defined probabilistic structure \cite{DBLP:journals/siamfm/ContL13,
DBLP:journals/economics/TokeP12, DBLP:journals/ior/ContST10}. For
instance,  Cont et al. \cite{DBLP:journals/ior/ContST10} introduced
a continuous-time stochastic model that presumes events such as
order arrivals and cancellations follow independent Poisson
processes, conditioned on the LOB's current state. Toke and Pomponio
\cite{DBLP:journals/economics/TokeP12} shown that a simple bivariate
Hawkes process fits nicely their empirical observations of
trades-through in limit order book. Additionally, Vvedenskaya et al.
\cite{DBLP:conf/isit/VvedenskayaSB11} formulated LOB dynamics as a
discrete-time Markov process governed by nonlinear ordinary
differential equations (ODEs). These stochastic models assume that
the LOB event stream follows strong probabilistic assumptions, but
they may fail to capture the evolution of the event stream due to
the variability in high-frequency markets. In comparison,
equilibrium models take a game-theoretic approach, focusing on
trader interactions and strategic behavior. These models use
subjective utility functions to represent payoffs associated with
different trading strategies \cite{DBLP:journals/rofs/Rocsu09,
DBLP:journals/rofs/Parlour98}. A prominent variant within
equilibrium models is the agent-based model (ABM)
\cite{DBLP:journals/anor/McGroartyBGC19}, where heterogeneous agents
with distinct behavior patterns interact within the LOB environment.
These methods struggle in highly noisy environments due to the
complexity introduced by agent parameters.

In recent years, deep learning has emerged as a powerful tool for
modeling and exploiting the dynamics of LOB. Shi and Cartlidge
\cite{DBLP:conf/kdd/ShiC22} combined stochastic point processes with
neural networks to model event stream patterns and proposed
PCT-LSTM, a state-dependent parallel neural Hawkes process for
predicting LOB events. Additionally, they also explored the
application of several neural network-based point processes to LOB
event stream prediction, such as self-attentive Hawkes process
(SAHP) \cite{DBLP:conf/icml/ZhangLKY20} and continuous-time LSTM
(CT-LSTM) \cite{DBLP:conf/nips/MeiE17}. Other studies, such as
DeepLOB \cite{DBLP:journals/tsp/ZhangZR19}, have further explored
feature engineering for LOB data, using LOB dynamics to predict
price trends. Generative adversarial networks (GANs) have been used
to replicate latent patterns within real order stream data
\cite{DBLP:conf/icaif/ColettaPCMBMVB21}. Another work, called
stock-GAN, employed GANs to generate realistic order streams that
capture several stylized facts observed in LOB data
\cite{DBLP:conf/aaai/LiWLSW20}. The LOB recreation model (LOBRM)
introduced by \cite{DBLP:conf/pkdd/ShiC21, DBLP:conf/aaai/ShiCC21}
employed continuous recurrent neural networks to predict volume
information at deeper price levels based on trade and quote data.

\textbf{Discussion.} The traditional approaches to modeling LOB
event steam often restrict it to specific stochastic processes or
lack support for advanced networks, limiting their effectiveness in
capturing the market's complex dynamics. Agent-based models and
recent deep learning-based methods, which use multiple agents to
simulate interactions among market participants or employ advanced
neural networks, improve the ability to capture event stream
dynamics. However, they are unable to achieve closed-form sampling,
which introduces challenges in forecasting LOB event stream. In
Table~\ref{tab:methods_comp}, we present a comparison of different
methods for event stream modeling to more clearly highlight their
distinctions.

\begin{table}[t]
\centering
\caption{Comparison of the proposed model with other LOB modeling approaches in terms of key properties.}
\label{tab:methods_comp}
\begin{threeparttable}
\resizebox{0.48\textwidth}{!}{ 
\begin{tabular}{|l|c|c|c|c|}
\hline \textbf{Model} & \textbf{No Restr.\(^{(1)}\)} & \textbf{No
Asmp.\(^{(2)}\)} & \textbf{Flexible\(^{(3)}\)} & \textbf{Closed-form
Sampling\(^{(4)}\)} \\ \hline Poisson
\cite{DBLP:journals/ior/ContST10}             &
\textcolor{red}{\ding{55}} & \textcolor{red}{\ding{55}} &
\textcolor{red}{\ding{55}} & \textcolor{green}{\ding{51}} \\ \hline
Hawkes \cite{DBLP:journals/economics/TokeP12}    & \textcolor{red}{\ding{55}} & \textcolor{green}{\ding{51}} & \textcolor{red}{\ding{55}} & \textcolor{red}{\ding{55}} \\
\hline
ODEs \cite{DBLP:conf/isit/VvedenskayaSB11} & \textcolor{green}{\ding{51}} & \textcolor{green}{\ding{51}}  & \textcolor{red}{\ding{55}} & \textcolor{green}{\ding{51}} \\ \hline
LOBRM \cite{DBLP:conf/aaai/ShiCC21}          & \textcolor{green}{\ding{51}} & \textcolor{green}{\ding{51}} & \textcolor{red}{\ding{55}} & \textcolor{red}{\ding{55}} \\ \hline
ABM \cite{DBLP:journals/anor/McGroartyBGC19}               & \textcolor{green}{\ding{51}} & \textcolor{green}{\ding{51}} & \textcolor{green}{\ding{51}} & \textcolor{red}{\ding{55}} \\ \hline
Stock-GAN \cite{DBLP:conf/aaai/LiWLSW20}               & \textcolor{green}{\ding{51}} & \textcolor{green}{\ding{51}} & \textcolor{green}{\ding{51}} & \textcolor{red}{\ding{55}} \\ \hline
GANs \cite{DBLP:conf/icaif/ColettaPCMBMVB21}                & \textcolor{green}{\ding{51}} & \textcolor{green}{\ding{51}} & \textcolor{green}{\ding{51}} & \textcolor{red}{\ding{55}} \\ \hline
DeepLOB \cite{DBLP:journals/tsp/ZhangZR19}         & \textcolor{green}{\ding{51}} & \textcolor{green}{\ding{51}} & \textcolor{green}{\ding{51}} & \textcolor{red}{\ding{55}} \\ \hline
PCT-LSTM \cite{DBLP:conf/kdd/ShiC22}           & \textcolor{red}{\ding{55}} & \textcolor{green}{\ding{51}} & \textcolor{green}{\ding{51}} & \textcolor{red}{\ding{55}} \\ \hline
\textbf{LOBDIF (ours)}   & \textcolor{green}{\ding{51}} & \textcolor{green}{\ding{51}} & \textcolor{green}{\ding{51}} & \textcolor{green}{\ding{51}} \\ \hline
\end{tabular}
} 
\begin{tablenotes}
\small
\item \(^{(1)}\) \parbox[t]{0.45\textwidth}{Without restriction on specific stochastic process.}
\item \(^{(2)}\) \parbox[t]{0.45\textwidth}{Without temporal dependence assumption.}
\item \(^{(3)}\) \parbox[t]{0.45\textwidth}{Any advanced network architecture can be utilized during the computation.}
\item \(^{(4)}\) \parbox[t]{0.45\textwidth}{Sampling without any approximation.}
\end{tablenotes}
\end{threeparttable}
\end{table}

\subsection{Diffusion Model}

Diffusion models have recently emerged as a powerful generative
modeling approach, demonstrating success across diverse application
domains, including image generation
\cite{DBLP:conf/cvpr/RombachBLEO22, DBLP:journals/tkde/XiaoHLLL23},
time series prediction and imputation \cite{DBLP:conf/icml/ShenK23,
DBLP:conf/cikm/KoaMNC23, DBLP:conf/icde/LiuHFS0F23,
DBLP:journals/pvldb/ChenZMLDLHRLZ23, DBLP:journals/tkde/LiuWLYWY24},
and data synthesis \cite{DBLP:journals/tkde/CaoTGXCHL24,
DBLP:journals/pacmmod/LiuFTLD24}. Here, the related work focuses on
the application of diffusion models to time series tasks, as LOB is
essentially a time series problem. These diffusion-based models have
demonstrated significant potential in various time series tasks,
including forecasting, generation, and imputation. Their ability to
capture complex spatio-temporal distribution and handle uncertainty
makes them well-suited for both univariate and multivariate time
series data.

Diffusion models are increasingly being applied to time series
forecasting, particularly for capturing stochastic properties within
the data. For instance, D-Va \cite{DBLP:conf/cikm/KoaMNC23} combined
deep hierarchical variational autoencoders (VAE) with diffusion
probabilistic methods to model stock price volatility, leading to
improved forecasting performance. Similarly, TimeDiff
\cite{DBLP:conf/icml/ShenK23} incorporated innovations such as
future mixup and autoregressive initialization, enhancing the
model's ability to capture intricate sequential patterns and improve
forecasting accuracy. In generative tasks, diffusion models have
proven valuable for time series applications. DOSE
\cite{DBLP:conf/nips/TaiL0TZ23} integrated diffusion models with
speech enhancement, providing a model-agnostic approach that
conditions the diffusion process with additional context to generate
clearer speech signals. Disffsformer
\cite{DBLP:journals/corr/abs-2402-06656} proposed a conditional
diffusion Transformer framework for stock forecasting, which
augments time-series stock data with label and industry information.
This demonstrates the potential of diffusion models in data
generation tasks beyond traditional predictive applications.
Additionally, diffusion models have been applied to imputing missing
values in time series data. PriSTI \cite{DBLP:conf/icde/LiuHFS0F23}
used a conditional feature extraction module based on linear
interpolation to handle missing data in spatio-temporal scenarios,
while ImDiffusion \cite{DBLP:journals/pvldb/ChenZMLDLHRLZ23} focused
on multivariate time series anomaly detection, applying a
conditional weight-incremental diffusion model to enhance imputation
and identify anomalies in time series data.

Although diffusion models have shown impressive performance across
various fields, their application in modeling the LOB event stream
remains relatively unexplored. This work explores the application of
diffusion models to LOB event stream prediction for the first time,
aiming to extend their applicability to this domain.

\section{Preliminaries}
\label{sec:define}

\begin{figure*}[t]
\centering
\includegraphics[width=0.85\textwidth]{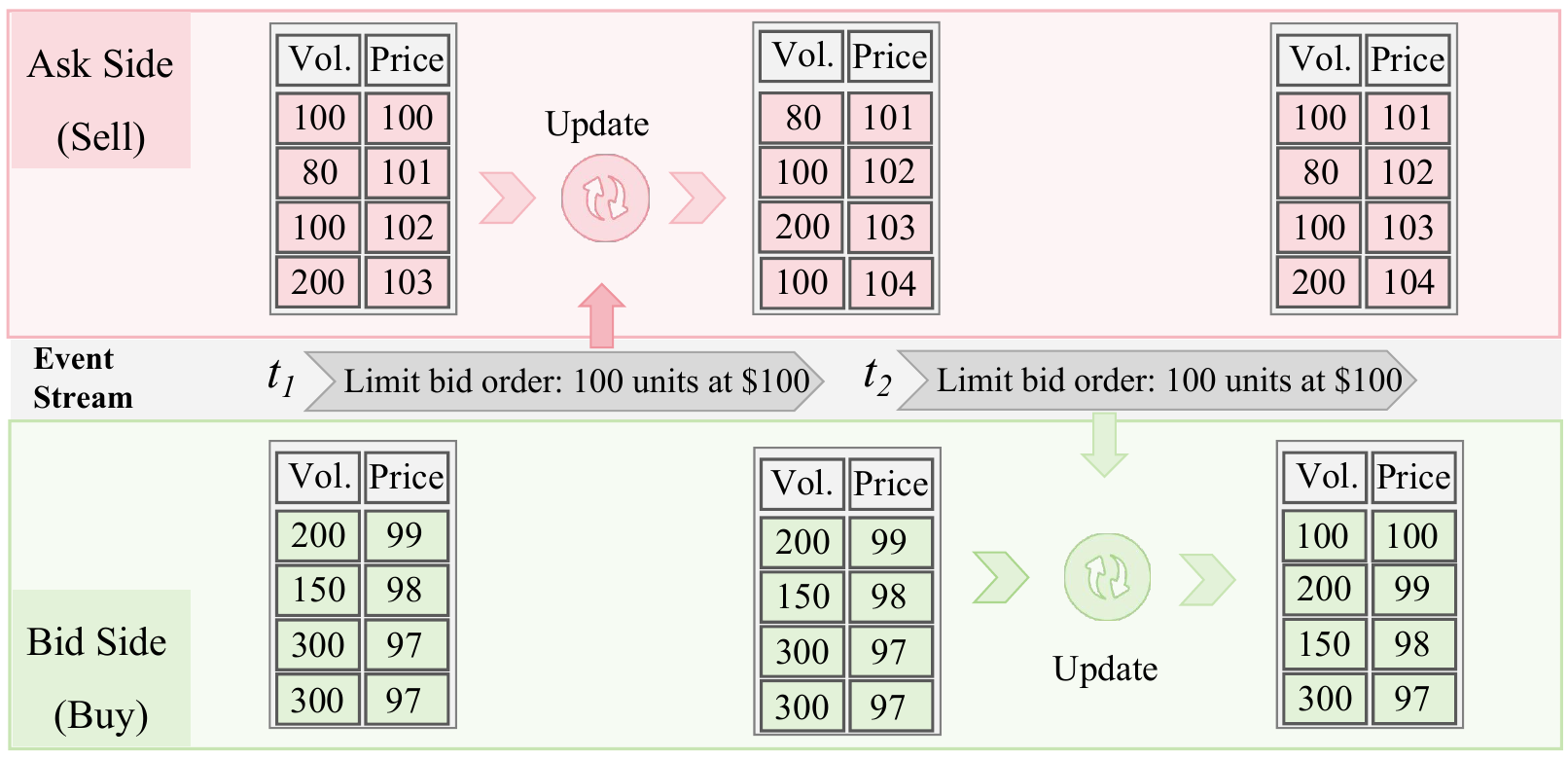}
\caption{An LOB with four price levels evolving over time. The event
at $t_1$ matches an order on the ask side, resulting in a successful
transaction, while the event at $t_2$ cannot find a suitable
matching price on the ask side and is instead recorded on the bid
side.} \label{fig:lob_demo}
\end{figure*}

\subsection{Limit Order Book}

Nearly all financial markets operate on the continuous double
auction mechanism \cite{DBLP:book/Friedman18}. Traders submit orders
specifying the maximum price they are willing to pay for a specified
quantity of an asset, or the minimum price they are willing to
accept to sell a quantity of the asset. An LOB is a central
component of modern financial markets that records submitted but
unexecuted orders. It provides a real-time, dynamic record of all
outstanding buy and sell orders at various price levels, effectively
serving as a mechanism for matching buy and sell orders. The LOB is
organized into two parts:
\begin{itemize}
    \item Bid side: The buy orders, representing participants willing to
purchase the asset at various prices.
    \item Ask side: The sell orders, where participants offer the asset for
sale at different price levels.
\end{itemize}

Figure~\ref{fig:lob_demo} presents a schematic of an LOB with 4
price levels evolving over time. Initially, the best bid is at a
price of $\$99$ with a volume of $200$, and the best ask is at a
price of $\$100$ with a volume of $100$. The next event (at time
$t_1$) is a new bid (buy) order at a price of $\$100$ with a volume
of $100$. This order executes in full against the best ask, which
causes the best ask price to rise to $\$101$ (the second-best ask
price has now become the best bid price). The following event (at
time $t_2$) is another bid at a price of $\$100$ and a volume of
$100$. However, this ask price is too low to execute (since the
current lowest ask is $\$101$), so the bid order remains unexecuted
and is recorded in the LOB on the bid side at the best bid price. As
the LOB provides the most detailed demand and supply information
available in the market, it is considered the ultimate microscopic
level of market structure \cite{DBLP:journals/qf/BouchaudMP02}. The
event stream within the LOB is closely tied to its evolution, making
the prediction of event streams crucial for understanding the
microscopic dynamics of the market.

\textbf{Event stream prediction}: Consider an event sequence
$\mathbf{S} = \{(t_1, e_1), \dots, (t_T, e_T)\} $, where $t_i \in
\mathbb{R}^+ $ represents the time and $e_i \in \{0, 1, 2, 3\}$
represents the event type of the $i$-th arrival in the sequence.
Event arrivals are categorized into four classes: submission of an
order at the bid side ($e=0$), cancellation of an order at the bid
side ($e=1$), submission of an order at the ask side ($e=2$), and
cancellation of an order at the ask side ($e=3$). The event stream
$\mathbf{S}$ is manually divided into equal-length $L$-sized
sub-streams using a rolling window approach with a step size of one,
denoted as $ \{\mathbf{X}_1, \mathbf{X}_2, \cdots,
\mathbf{X}_{T-L+1} \}$. Given a sub-stream $\mathbf{X}_{j \in
\{1,\cdots,T-L+1\}} =\{\mathbf{x}_1, \cdots, \mathbf{x}_L \} =
\{(t_1, e_1), \dots, (t_L, e_L)\}$ of length $L$, the model receives
this event sub-stream as input and makes a prediction for the next
event's occurrence time $t_{L+1}$ and event type $e_{L+1}$.

\begin{figure*}[t]
\centering
\includegraphics[width=0.95\textwidth]{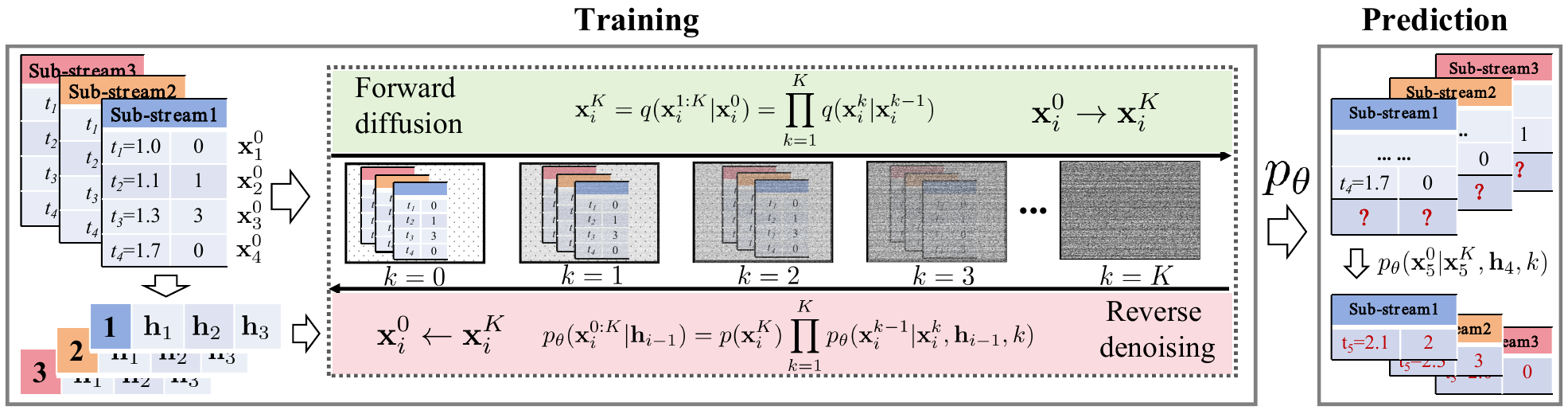}
\caption{The overall framework of our proposed method.}
\label{fig:framework}
\end{figure*}

\subsection{Diffusion Model}

Inspired by non-equilibrium thermodynamics, the diffusion model
defines a Markov chain of forward diffusion steps that gradually add
random noise to data, followed by a reverse denoising process that
learns to reconstruct the desired data samples from Gaussian noise.
Specifically, given a data point sampled from a real data
distribution $\mathbf{x}^0\sim q(\mathbf{X}_j)$ (the initial
noise-free observation), the forward diffusion process incrementally
adds small amounts of Gaussian noise
$\{\mathbf{\epsilon}^k\}_{k=1}^K$ over $K$ steps, resulting in a
sequence of progressively noisier samples
\{$\mathbf{x}^1,\ldots,\mathbf{x}^K\}$. The noise levels at each
forward diffusion step are controlled by a variance schedule
$\{\beta_k\in(0,1)\}_{k=1}^K$. Specifically, at step $k$, the
observation $\mathbf{x}^k$ can be obtained from the previous
observation $\mathbf{x}^{k-1}$ as follows: $\mathbf{x}^k =
\sqrt{1-\beta_k}\mathbf{x}^{k-1} + \beta_k \mathbf{\epsilon}^k$,
where $\sqrt{1-\beta_k}\mathbf{x}^{k-1}$ represents the scaled
contribution of the previous observation, and $\beta_k
\mathbf{\epsilon}^k$ indicates the degree of added noise. Since
$\mathbf{x}^{k-1}$ can be inferred from $\mathbf{x}^0$, the
observation $\mathbf{x}^{k}$ can be expressed in a closed-form as:
$\mathbf{x}^k = \sqrt{\overline{\alpha}_k} \mathbf{x}^0 + (1 -
\overline{\alpha}_k) \mathbf{\epsilon}^k$, where
$\overline{\alpha}_k = \prod_{k=1}^K \alpha_k$ and $\alpha_k = 1 -
\beta_k$. $\mathbf{\epsilon}^k$ is the sampled Gaussian noise,
denoted as $\mathbf{\epsilon}^k \sim \mathcal{N}(0,
\boldsymbol{I})$. As the step $k$ increases, the sampled data
$\mathbf{x}^0$ gradually loses its distinguishable features.
Eventually, as $K \to \infty$, $\mathbf{x}^K$ converges to a
Gaussian distribution.

Mathematically, the entire forward diffusion process is defined as
follows:
\begin{equation}
\begin{aligned}
\mathbf{x}^K = q(\mathbf{x}^{1:K}|\mathbf{x}^0)&=\prod_{k=1}^Kq(\mathbf{x}^k|\mathbf{x}^{k-1}), \\
\mathbf{x}^k = q(\mathbf{x}^k|\mathbf{x}^{k-1})&=\sqrt{1-\beta_k}\mathbf{x}^{k-1} + \beta_k \mathbf{\epsilon}^k\\
&=\sqrt{\overline{\alpha}_k} \mathbf{x}^0 + (1 - \overline{\alpha}_k) \mathbf{\epsilon}^k.
\end{aligned}
\label{eq:diffusion_forward}
\end{equation}

Conversely, the reverse denoising process aims to recover
$\mathbf{x}^0$ starting from $\mathbf{x}^K$, where $\mathbf{x}^K
\sim \mathcal{N}(0, \boldsymbol{I})$. The observation $\mathbf{x}^0$
can be inferred progressively from $\mathbf{x}^K$ as follows:
\begin{equation}
    \begin{aligned}        \mathbf{x}^0&=p_\theta(\mathbf{x}^{0:K})=p(\mathbf{x}^K)\prod_{k=1}^K p_\theta(\mathbf{x}^{k-1}|\mathbf{x}^k)\:,\\
    \mathbf{x}^{k-1}&=p_\theta(\mathbf{x}^{k-1}|\mathbf{x}^k)\\
    &=\mathcal{N}(\mathbf{x}^{k-1}; \frac1{\sqrt{\alpha_k}}\Big(\mathbf{x}^k-\frac{1-\alpha_k}{\sqrt{1-\bar{\alpha}_k}}\mathbf{\epsilon}^k\Big),  \sigma^2)\\
    &= \frac1{\sqrt{\alpha_k}}\Big(\mathbf{x}^k-\frac{1-\alpha_k}{\sqrt{1-\bar{\alpha}_k}}\mathbf{\epsilon}_\theta(\cdot)\Big) +   \sigma \mathbf{\epsilon}, \mathbf{\epsilon} \sim \mathcal{N}(0, \boldsymbol{I}),
    \end{aligned}
    \label{eq:reverse}
\end{equation}
where $\sigma =
\frac{1-\bar{\alpha}_{k-1}}{1-\bar{\alpha}_k}\cdot\beta_k$ and
$p_\theta(\mathbf{x}^{k-1}|\mathbf{x}^k)$ is derived from
conditional Bayesian inference. $\mathbf{\epsilon}^k$ is
unknown during the denoising process and must be estimated by the
neural network. The key idea of the denoising process is to use a
neural network $\mathbf{\epsilon}_\theta(\cdot)$ parameterized by
$\theta$, which takes $\mathbf{x}^k$ and $k$ as inputs and outputs
the estimated noise $\mathbf{\epsilon}^k$. The neural network
$\mathbf{\epsilon}_\theta(\cdot)$ can be effectively optimized with
the following simplified mean squared error (MSE) objective
\cite{DBLP:conf/nips/HoJA20}:
\begin{equation}
    \begin{aligned}
        \mathcal{L}_k& =\mathbb{E}_{k\sim[1,K], \mathbf{x}^0, \mathbf{\epsilon}^k}\left[\|\mathbf{\epsilon}^k-\mathbf{\epsilon}_\theta(\mathbf{x}^k,k)\|^2\right] \\
        &=\mathbb{E}_{k\sim[1,K], \mathbf{x}^0, \mathbf{\epsilon}^k}\left[\left\|\mathbf{\epsilon}^k-\mathbf{\epsilon}_\theta(\sqrt{\bar{\alpha}_k}\mathbf{x}^0+\sqrt{1-\bar{\alpha}_k}\mathbf{\epsilon}^k,k)\right\|^2\right].
    \end{aligned}
    \label{eq:loss}
\end{equation}
The objective of the above function is to guide the neural network
$\mathbf{\epsilon}_\theta(\cdot)$ in estimating the Gaussian noise
added to the input $\mathbf{x}^k$ and to minimize the error between
the actual noise and the predicted noise. Once
$\epsilon_\theta(\cdot)$ can accurately predict the noise, it can
transform $\mathbf{x}^k$ to $\mathbf{x}^{k-1}$. This enables the
progressive recovery of $\mathbf{x}^0$ from the initially sampled
noise $\mathbf{x}^K$.

As described in Eq.~\eqref{eq:reverse}, each step of the reverse
process is explicitly modeled using a deterministic function that
incorporates $ \mathbf{x}^k$, the predicted noise from
$\epsilon_\theta(\cdot)$, and an additional Gaussian noise
$\mathbf{\epsilon}$. These components enable the diffusion model to
inherently support \emph{closed-form sampling}, as it eliminates the
need for computationally expensive operations such as Markov chain
Monte Carlo or rejection sampling to estimate these variables.

The preliminary information above offers a foundational
understanding of the limit order book and diffusion model. Next, we
will provide a detailed explanation of how to apply the diffusion
model to predict event stream in the limit order book.

\section{LOBDIF}
\label{sec:method}

Figure~\ref{fig:framework} illustrates the overall framework of
LOBDIF, which consists of a forward diffusion process and a reverse
denoising process. In the forward diffusion process, noise is
gradually added to the time-event samples. In the reverse denoising
process, the noise is progressively removed to recover the original
sample. Given the complex relationship between time and event types
in LOB event stream prediction, a time-event encoder is employed to
better capture the representations of time and event types, along
with a carefully designed denoising network that leverages these
representations for effective reverse denoising.

\subsection{Forward Diffusion Process}
\label{sec:forward}

For each event $\mathbf{x}_i^0 = (t_i, e_i)$ in the sub-stream, we
model the forward process as a Markov chain. Starting from
$\mathbf{x}_i^0 = (t_i, e_i)$, we progressively add small amounts of
Gaussian noise $\mathbf{\epsilon}_i^1$ to both the time and event
components, producing $\mathbf{x}_i^1$. This noise-adding process is
repeated iteratively until step $K$ is reached, at which point
$\mathbf{x}_i^0$ evolves into $\mathbf{x}_i^K$, resulting in the
sequence $(\mathbf{x}_i^0, \mathbf{x}_i^1, \dots, \mathbf{x}_i^K)$.
The complete forward process for the event is defined as follows:
\begin{equation}
    \begin{aligned}
        &q(\mathbf{x}_i^k|\mathbf{x}_i^{k-1})=(q(t_i^k|t_i^{k-1}), q(e_i^k|e_i^{k-1})),\\
        &t^k_i=q(t_i^k|t_i^{k-1})= \sqrt{1-\beta_k}t_i^{k-1} + \beta_k\mathbf{\epsilon}_i^k,\\
        &e^k_i=q(e_i^k|e_i^{k-1})= \sqrt{1-\beta_k}e_i^{k-1} + \beta_k\mathbf{\epsilon}_i^k.\\
    \end{aligned}
\end{equation}
This forward diffusion process is similar to the image-based
diffusion models described in Eq.~\eqref{eq:diffusion_forward}, with
the key difference being that, while image-based models add Gaussian
noise at the pixel level, our approach adds Gaussian noise to the
time $t$ and event type $e$ in the sampled points.

\subsection{Reverse Denoising Process}
\label{sec:denoise}

The reverse denoising process aims to iteratively reconstruct the
point $\mathbf{x}_i^0=(t_i,e_i)$ from $\mathbf{x}_i^K \sim
\mathcal{N}(0, \boldsymbol{I})$. Unlike the image-based denoising
process described in Eq.~\eqref{eq:reverse}, event stream is
inherently sequential data. This means that when predicting each
step $\mathbf{x}_i^0$, we not only rely on $\mathbf{x}_i^K$ and the
step $k$ as in image-based diffusion models, but also need to
account for the historical information prior to $i$. Such a setup
introduces more complexity, requiring the diffusion model to
consider additional dependencies during the denoising process. To
account for these dependencies, we encode the previous events,
$\{(t_1, e_1), \cdots, (t_{i-1}, e_{i-1})\}$, into a historical
context $\mathbf{h}_{i-1}$, which then serves as the condition for
predicting the next event $(t_i, e_i)$. This can be formulated as
follows:
\begin{equation}
    \begin{aligned}
    p_\theta(\mathbf{x}_i^{0:K}|\mathbf{h}_{i-1})=&p(\mathbf{x}_i^K)\prod_{k=1}^Kp_\theta(\mathbf{x}_i^{k-1}|\mathbf{x}_i^k, \mathbf{h}_{i-1}, k),\\
    \small
    p_\theta(\mathbf{x}_i^{k-1}|\mathbf{x}_i^k,\mathbf{h}_{i-1}, k)=&p_\theta(t_i^{k-1}|t_i^k, e_i^k, \mathbf{h}_{i-1}, k) \cdot \\ &p_\theta(e_i^{k-1}|t_i^k, e_i^k, \mathbf{h}_{i-1}, k),
    \end{aligned}
    \label{eq:lobdif_reverse}
\end{equation}
where $p(\mathbf{x}_i^K)$ represents sampling a noise from a
Gaussian distribution. The key component in the conditional
probability function $p_\theta(\cdot|\cdot)$ is the neural network
$\mathbf{\epsilon}_\theta$, which predicts the noise
$\mathbf{\epsilon}_i^k$ based on the current time $t_i^k$, event
type $e_i^k$, denoising step $k$ and historical context information
$\mathbf{h}_{i-1}$. With the predicted noise, it becomes possible to
derive time and event type predictions according to
Eq.~\eqref{eq:reverse}. This approach allows us to decompose the
modeling of the joint time-event distribution into conditionally
independent components, which facilitates more effective modeling of
the observed time-event distribution.

Next, we will explain how to obtain the historical context
$\mathbf{h}_{i-1}$ through time-event encoding and how to design the
denoising network to compute the previous time $t_i^{k-1}$ and event
type $e_i^{k-1}$ through denoising network.

\subsubsection{Time-event Encoding}

Given the irregular arrival time of events and the discrete nature
of their types, we need to effectively represent both time and event
types to facilitate capturing the underlying patterns in the order
stream. For the time representation, we use a positional encoding
method, which encodes time as a continuous representation, as
described in Eq.~\eqref{eq:position_encod}. This approach is
well-suited for our context because it allows the model to learn
relative time relationships between events.
\begin{equation}
    \phi(t)[j] =
    \begin{cases}
    \cos\left( \frac{t}{10000}^{\frac{j-1}{M}} \right) & \text{if } j \text{ is odd} \\
    \sin\left( \frac{t}{10000}^{\frac{j-1}{M}} \right) & \text{if } j \text{ is even}
    \end{cases},
    \label{eq:position_encod}
\end{equation}
where $\phi(t)$ denotes the temporal embedding and $M$ is the
embedding dimension. For the event type, we first encode it as a
one-hot vector. To obtain a more expressive representation, we then
apply a linear transformation to the one-hot vector to obtain a
continuous representation $\phi(e)$ with $M$ dimensions. In this
way, both time and event types are represented as continuous vectors
with the same dimensionality.

For each event $\mathbf{x} = (t, e)$, we obtain the event-temporal
embedding $\phi(t, e)$ by adding the time encoding $\phi(t)$ and the
event embedding $\phi(e)$ element-wise. The embedding for the entire
event sequence $\mathbf{X} = \{(t_i, e_i)\}_{i=1}^{L} \}$ is then
represented as $\Phi(t, e) = \{\phi_1, \phi_2, \dots, \phi_L\} \in
\mathbb{R}^{L \times M}$, where each $\phi_i = \{\phi(t, e)_i\}$.
Additionally, we maintain the separate temporal embedding $\Phi(t) =
\{\phi(t)_1, \phi(t)_2, \dots, \phi(t)_L\}$ and event embedding
$\Phi(e) = \{\phi(e)_1, \phi(e)_2, \dots, \phi(e)_L\}$. These
embeddings are designed to capture distinct characteristics of the
temporal and event aspects, allowing the model to adaptively
leverage these representations for improved performance.

After the initial event embedding and temporal encoding layers, we
pass $\Phi(t, e)$, $\Phi(t)$ and $\Phi(e)$ through three
self-attention modules. Specifically, the scaled dot-product
attention \cite{DBLP:conf/nips/VaswaniSPUJGKP17} is defined as:
\begin{equation}
    \begin{aligned}
        &\text{Attention}(Q,K,V)=\text{Softmax}(\frac{QK^T}{\sqrt{d}}),\\
        &S=\text{Attention}(Q, K, V) V,
    \end{aligned}
\end{equation}
where $Q$, $K$, and $V$ represent the queries, keys, and values,
respectively. Taking $\Phi(t, e)$ as an example, the self-attention
operation first takes the embedding $\Phi(t, e)$ as input, which is
then transformed into three matrices via linear projections: $Q =
\Phi(t, e) W^Q, K = \Phi(t, e) W^K, V = \Phi(t, e) W^V$, where
$W^Q$, $ W^K$, and $W^V$ are the weight matrices for the respective
linear projections. Afterward, a position-wise feed-forward network
is applied to the attention output $S$ to produce the hidden
representation $h(t, e)$. For the other two embeddings, $\Phi(t)$
and $\Phi(e)$, the same self-attention operation is used to generate
the hidden temporal representation $h(t)$ and event representation
$h(e)$. The final time-event representation of sequence
$\mathbf{X}$, denoted as $\mathbf{h}$, is the concatenation of these
three representations, i.e., $\mathbf{h} = [h(t, e) \parallel h(t)
\parallel h(e)] \in \mathbb{R}^{L \times (3\times M)}$.
Additionally, $\mathbf{h}_{i-1}$ refers to the representation at the
$i-1$ position of $\mathbf{h}$.

\subsubsection{Denoising Network}

We design a denoising network to capture the interdependence between
time and event, which facilitates the learning of time-event joint
distributions. Specifically, it performs time and event attentions
simultaneously at each denoising step to capture fine-grained
relations. Each step of the denoising process shares the same
structure, which takes in the previously predicted values
$\mathbf{x}^{k+1}_i = (e_{i}^{k+1}, t_i^{k+1})$, denoising step $k$
and the history context representation $\mathbf{h}_{i-1}$ to achieve
conditional denoising. We begin by representing the denoising step
$k$ as a vector to facilitate the model's processing, using
Sinusoidal positional embedding
\cite{DBLP:conf/nips/VaswaniSPUJGKP17} to encode $k$. Next, we
simultaneously compute the time attention $\omega_{t}$ and event
attention $\omega_{e}$ based on the condition $\mathbf{h}_{i-1}$ and
the current denoising step $k$. The goal of the time attention and
event attention is to generate context vectors by attending to
specific parts of the time input and event type input, respectively.
The introduction of $\mathbf{h}_{i-1}$ allows us to account for the
dynamic nature of the time-event relationships in the event stream,
which impacts the prediction of the next event $\mathbf{x}_i^0$. The
inclusion of $k$ considers the dynamics of the diffusion process.
These two elements enable the model to adjust its attention based on
the evolving dynamics of both time-event relationships and the
diffusion process, thereby capturing the complexity of the
time-event sequence. The mathematical representation of this process
is given as follows:
\begin{equation}
    \begin{aligned}
        &\phi_{k}=\text{PositionEmb}(k),\\
        &\omega_{t}=\mathrm{Softmax}(f([\mathbf{h}_{i-1} \parallel \phi_{k}])),\\
        &\omega_{e}=\mathrm{Softmax}(f([\mathbf{h}_{i-1} \parallel \phi_{k}])),
    \end{aligned}
\end{equation}
where $f$ denotes the feed-forward network and $\omega_{t}$ and
$\omega_{e}$ measure the mutual dependence between time and event.
We then combine the historical context of time and event, $h(t)_{i-1}$ and
$h(e)_{i-1}$, the previously predicted values $\mathbf{x}^{k+1}_i =
(e_{i}^{k+1}, t_i^{k+1})$, along with the denoising step $k$, into a
feed-forward neural network to model the evolution of time and
events. Each layer is formulated as follows:
\begin{equation}
    \begin{aligned}
        &\hat{t}^k_i=\mathrm{ReLU}(f([t^{k+1}_i + h(t)_{i-1} + \phi_{k}])),\\
        &\hat{e}^k_i=\mathrm{ReLU}(f([e_{i}^{k+1} + h(e)_{i-1} + \phi_{k}])),\\
    \end{aligned}
\end{equation}
where $\mathrm{ReLU}$ denotes the activation function. Finally, the
predicted noise is generated by combining time attention, event
attention, and the predicted values $\mathbf{x}_i^k$, and is given
by the following output:
\begin{equation}
    \begin{aligned}
        &\hat{\mathbf{x}}_{i}^k=[\hat{t}^k_i, \hat{e}^k_i],\\
        &\epsilon_{t, i}^{k}=\sum\omega_{t}\hat{\mathbf{x}}_{i}^k, \epsilon_{e, i}^{k}=\sum\omega_{e}\hat{\mathbf{x}}_{i}^k,
    \end{aligned}
\end{equation}
where $\epsilon_{t,i}^{k}$ and $\epsilon_{e,i}^{k}$ are the
predicted noise at step $k$ for the $i$-th event. With the predicted
noise $\mathbf{\epsilon}_{i}^{k}=(\epsilon_{t,i}^{k},
\epsilon_{e,i}^{k})$, we can optimize $\theta$ with the true noise
$\mathbf{\epsilon}_i^k$ added in the forward process. In this way,
the interdependence between time and event is captured adaptively
and dynamically, facilitating the learning of the time-event joint
distribution.

\section{Training and Prediction}
\label{sec:train_infer}

\subsection{Training}
\label{sec:training}

The training of LOBDIF is based on a similar derivation as in
Eq.~\eqref{eq:loss}, with the key difference being that the neural
network $\mathbf{\epsilon}_{\theta}(\cdot)$ additionally receives
the historical context $\mathbf{h}_{i-1}$, which accounts for the
history of events and their types. The final loss function of LOBDIF
is shown as follows:
\begin{equation}
    \begin{aligned}
        &\mathcal{L}_k =\mathbb{E}_{k\sim[1,K], \mathbf{x}_i^0, \mathbf{\epsilon}_i^k}\left[\|\mathbf{\epsilon}_i^k-\mathbf{\epsilon}_\theta(\mathbf{x}_i^k, \mathbf{h}_{i-1}, k)\|^2\right] \\
        &=\mathbb{E}_{k\sim[1,K], \mathbf{x}_i^0, \mathbf{\epsilon}_i^k}\left[\left\|\mathbf{\epsilon}_i^k-\mathbf{\epsilon}_\theta(\sqrt{\bar{\alpha}_k}\mathbf{x}_i^0+\sqrt{1-\bar{\alpha}_k}\mathbf{\epsilon}_i^k, \mathbf{h}_{i-1}, k)\right\|^2\right].
    \end{aligned}
    \label{eq:loss2}
\end{equation}

For each event $\mathbf{x}_i=(t, e)$ in the sequence $\mathbf{X}$, the
forward diffusion process will be executed for $K$ steps, generating
$K$ observations $\{\mathbf{x}_i^{k}\}_{k=1}^K$. Similarly,
events in all sequences $\mathbf{X}_{j \in \{1,\cdots,T-L+1\}}$ will
undergo the same forward diffusion process, producing a large number
of observations. All these observations are included in the training
set. The overall framework will be trained in an end-to-end manner.
The pseudocode for the training procedure is shown in
Algorithm~\ref{alg:training}.

\begin{algorithm}[t] \small
\caption{Training for each event $\mathbf{x}_i$}
\begin{algorithmic}[1]
\Require $\mathbf{h}_{i-1}$
\Repeat
    \State $\mathbf{x}_i^0 \sim q(\mathbf{X})$
    \State $k \sim \text{Uniform}(1, 2, \dots, K)$
    \State $\mathbf{\epsilon}_i^k \sim \mathcal{N}(0, I)$
    \State Take gradient descent step on
    \[
    \nabla_{\theta} \| \mathbf{\epsilon}_i^k- \epsilon_{\theta}(\sqrt{\bar{\alpha}_k} \mathbf{x}_i^0 + \sqrt{1 - \bar{\alpha}_k} \mathbf{\epsilon}_i^k, \mathbf{h}_{i-1}, k) \|^2
    \]
\Until{Converged}
\end{algorithmic}
\label{alg:training}
\end{algorithm}

\subsection{Prediction with Skip-step Sampling}

To predict the event $\mathbf{x}_{i}=(t_{i}, e_{i})$ at timestep $i$
with a trained LOBDIF, we first obtain the hidden representation
$\mathbf{h}_{i-1}$ by employing the time-event encoder given past $i$ event
sequence. Then, we can predict the next event starting from Gaussian
noise $t_{i}^K, e_{i}^K \thicksim \mathcal{N}(0, \boldsymbol{I})$
conditioned on $\mathbf{h}_{i-1}$ based on Eq.~\eqref{eq:reverse}.
However, in practical LOB prediction scenarios, computational
efficiency is often strictly required, whereas the reverse denoising
process described in Eq.~\ref{eq:reverse} is a time-consuming
step-by-step procedure. To make LOBDIF suitable for LOB event stream
prediction task, inspired by image-based diffusion models
\cite{DBLP:conf/iclr/ZhangTC23}, we reformulate
$p_\theta(\mathbf{x}^{k-1}|\mathbf{x}^k)$ in Eq.~\eqref{eq:reverse}
using a reparameterization trick to enable skip-step sampling.

Assuming we start from step $k$ and skip to step $s$ (where $s < k$)
instead of performing step-by-step denoising, according to
Eq.~\eqref{eq:diffusion_forward}, we can define $\mathbf{x}_i^s$ as
Step 1 in Eq.~\eqref{eq:skip-step}.
\begin{equation}
    \small
    \begin{aligned}
    &\mathbf{x}_i^{s} =\sqrt{\bar{\alpha}_{s}} \cdot p(\mathbf{x}_i^0|\mathbf{x}_i^k)+\sqrt{1-\bar{\alpha}_{s}}\mathbf{\epsilon}_i^{s}, \quad Step 1 \\
    &=\sqrt{\bar{\alpha}_{s}} \cdot p(\mathbf{x}_i^0|\mathbf{x}_i^k) +\sqrt{1-\bar{\alpha}_{s}-\sigma_k^2}\mathbf{\epsilon}_\theta(\mathbf{x}_i^k, \mathbf{h}_{i-1}, k)+\sigma_k\mathbf{\epsilon}, \quad Step 2\\
    \end{aligned}
    \label{eq:skip-step}
\end{equation}
where
$p(\mathbf{x}_i^0|\mathbf{x}_i^k)=\frac{\mathbf{x}_i^k-\sqrt{1-\bar{\alpha}_k}\mathbf{\epsilon}_\theta(\mathbf{x}_i^k,
\mathbf{h}_{i-1}, k)}{\sqrt{\bar{\alpha}_k}}$, inferred from
Eq.~\eqref{eq:diffusion_forward} and
$\sigma_k=\frac{1-\bar{\alpha}_{k-1}}{1-\bar{\alpha}_k}\cdot\beta_k$.
We can observe that the equation in Step 1 is unsolvable because it
depends on $\mathbf{\epsilon}_i^s$, whose value is estimated based
on $\mathbf{x}_i^s$, the very variable we aim to solve. To address
this, we apply the reparameterization trick, replacing
$\sqrt{1-\bar{\alpha}_{s}}\mathbf{\epsilon}_i^{s}$ with
$\sqrt{1-\bar{\alpha}_{s}-\sigma_t^2}\mathbf{\epsilon}_i^k +
\sigma_t\mathbf{\epsilon}$, as described in Step 2. The
reparameterization preserves the mean and variance while enabling
$\mathbf{\epsilon}_i^s$ to be estimated through $\mathbf{x}_i^k$ and
sampled noise $\mathbf{\epsilon}$. Note that the initial value of
$k$ is $K$, with $\mathbf{x}_i^K$ being known. Given
$\mathbf{x}_i^K$, it is possible to progressively derive
$\mathbf{x}_i^s$, which serves as the starting point for the next
skip step.

The skip-step sampling strategy, an extension of the denoising
diffusion implicit models \cite{DBLP:conf/iclr/SongME21,
DBLP:conf/iclr/ZhangTC23} from image-based diffusion to the event
stream scenario, incorporates sequential attributes such as
historical context. A key advantage of this reparameterization is
that it eliminates the need for step-by-step sampling, allowing for
arbitrary step-size sampling (denoted as $\tau$) without the need of
model retraining. The detailed denoising process is outlined in
Algorithm~\ref{alg:skip-step}. This approach significantly enhances
the efficiency of the reverse denoising process, as demonstrated in
the following example.

\begin{algorithm}[t] \small
\caption{Predicting $\mathbf{x}_i^0$}
\begin{algorithmic}[1]
\Require Gaussian noise $\mathbf{x}_i^K \sim \mathcal{N}(0, I)$ and $\mathbf{h}_{i-1}$\\
\textbf{Initialize step pairs:} \\
\textit{pair}=$\{(K, K-\tau), (K-\tau, K-2*\tau), \cdots, (\tau,
0))\}$ \For{$(k, s)$ in \textit{pair}}
    \If{$k\geq \tau$}
    \State $\mathbf{\epsilon} \sim \mathcal{N}(0, I)$
    \Else
    \State $\mathbf{\epsilon} = 0$
    \State $\mathbf{x}_i^{s} = \sqrt{\bar{\alpha}_{s}} \cdot p(\mathbf{x}_i^0|\mathbf{x}_i^k) +\sqrt{1-\bar{\alpha}_{s}-\sigma_k^2}\mathbf{\epsilon}_\theta(\mathbf{x}_i^k, \mathbf{h}_{i-1}, k)+\sigma_k\mathbf{\epsilon}$
\EndIf \EndFor \State \Return $\mathbf{x}_i^0$
\end{algorithmic}
\label{alg:skip-step}
\end{algorithm}

\begin{example}
Assuming we progressively add $K = \{1, \cdots, 1000\}$ steps of
noise $\{\epsilon^1, \cdots, \epsilon^{1000}\}$ to $\mathbf{x}^0$
and then train $\mathbf{\epsilon}_\theta(\cdot)$ according to
Eq.~\eqref{eq:loss2}. Once $\mathbf{\epsilon}_\theta(\cdot)$ is
trained, the traditional inference process would first sample the
noise $\mathbf{x}^{1000}$, then progressively predict the
$\epsilon^k$ value at each step, and obtain $\mathbf{x}^{k}$ from
$\mathbf{x}^{1000}$, eventually reconstructing $\mathbf{x}^0$. With
skip-step sampling strategy, however, we do not need to follow the
entire chain but can instead sample from any subset of steps. For
example, we could follow the subset $\{100, 200, \cdots, 900,
1000\}$ rather than the entire chain $\{1, \cdots, 1000\}$, which
enables a 10$\times$ speed-up in the reverse denoising process.
\end{example}

\section{Experiments}
\label{sec:exp}

In this section, we conduct experiments to study the following
research questions:
\begin{itemize}
    \item \textbf{RQ1}: How does the model's performance compare with existing baseline methods? We present the comparison in Section~\ref{sec:exp.compare}.
    \item \textbf{RQ2}: How do the two key components---the time-event encoding and the denoising network---impact model performance? Additionally, how does the skip-step sampling strategy affect the prediction stage? The results are presented in Section~\ref{sec:exp.ablation}.
    \item \textbf{RQ3}: How do the key hyperparameters (such as the numbers of forward steps, training epochs, and encoding dimensions) affect the effectiveness of the model? The results are discussed in Section~\ref{sec:exp.hyper}.
    \item \textbf{RQ4}: How can we gain a better understanding of the reverse denoising diffusion process? The results are presented in Section~\ref{sec:exp.case}.
\end{itemize}

\subsection{Experimental Setup}
\label{sec:exp.setup}

\subsubsection{Datasets}

The six datasets used in our study are sourced from three different
markets and time periods. For both MSFT1 and MSFT2, the data are
obtained from the NASDAQ market via the LOBSTER
platform\footnote{https://lobsterdata.com/info/DataSamples.php}. We
select order data from two different days, dividing each day into
80\% for training, 10\% for validation, and 10\% for testing. The
second and third stocks are Pingan Bank from the Shenzhen stock
exchange and China Telecommute from the Shanghai stock exchange,
both sourced from the CSMAR
database\footnote{https://www.csmar.com}. Due to differences in
granularity between the two databases, the types of order stream
data provided also vary. The LOBSTER dataset includes four types of
orders: submit and cancel for both the bid side and the ask side. In
contrast, the CSMAR dataset offers only three types of orders: bid,
ask, and an unidentified order type. More detailed statistics about
the datasets are summarized in Table~\ref{tab:bench_data_stat}.

\begin{table}[t]
  \caption{Statistics of datasets.}
  \label{tab:bench_data_stat}
  \center
  \begin{tabular}{c@{\hskip 5pt}c@{\hskip 5pt}c@{\hskip 5pt}cc@{\hskip 2pt}c@{\hskip 2pt}}
    \toprule
    Datasets & Training events  & Valid events & Test events & Types \\
    \toprule
    MSFT-1& 343664  & 42957 & 42958 & 4 \\
    MSFT-2& 303350  & 37918 & 37918 & 4 \\
    \midrule
    Pingan-1& 77265 & 14499 & 14498 & 3 \\
    Pingan-2& 53417 & 9672 & 9669 & 3 \\
    \midrule
    Telecom-1& 80107 & 15033 & 15033 & 3 \\
    Telecom-2& 65126 & 10026 & 10031 & 3 \\
    \bottomrule
  \end{tabular}
\end{table}

\subsubsection{Evaluation Baselines}

We conduct a comprehensive comparison of the proposed LOBDIF against
various state-of-the-art models designed for predicting event
arrivals. The evaluation includes five models: (1) Hawkes: A
state-dependent stochastic Hawkes point process model with an
exponential decaying kernel
\cite{DBLP:journals/qf/Morariu-PatrichiP22}. This is a stochastic
probability-based model that does not rely on neural networks. We
implemented it using the open-source library
tick\footnote{https://x-datainitiative.github.io/tick/index.html},
providing a clear contrast to other neural network-based models. (2)
LSTM: A straightforward LSTM-based model that does not include event
intensity rate modeling \cite{DBLP:journals/neco/HochreiterS97},
providing a sharp contrast to stochastic probability-based models.
(3) CT-LSTM: A neural Hawkes process that utilizes a continuous-time
LSTM unit to model intensity rates for all event types
\cite{DBLP:conf/nips/MeiE17}. This model combines the strengths of
LSTM and Hawkes processes for event prediction. (4) SAHP: The
self-attentive Hawkes process, which replaces the recurrent input
structure with an attention mechanism for enhanced representation
learning \cite{DBLP:conf/icml/ZhangLKY20}. (5) PCT-LSTM: The
state-dependent neural Hawkes process employs stacked CT-LSTM units
to separately model the intensity rates for different types of
events \cite{DBLP:conf/kdd/ShiC22}. This model leverages an enhanced
LSTM architecture to simulate event intensities while incorporating
the market state for improved prediction accuracy.

\begin{table*}[t]
  \caption{Performance comparison of all models on 6 datasets.}
  \label{tab:bas_comp}
  \center
  \resizebox{1.0\textwidth}{!}{
  \begin{tabular}{c|cc|cc|cc|cc|cc|cc}
    \toprule
    \multirow{2}{*}{Datasets}& \multicolumn{2}{c}{MSFT1} &
    \multicolumn{2}{|c}{MSFT2} &
    \multicolumn{2}{|c}{PINGAN1}&
    \multicolumn{2}{|c}{PINGAN2}&
    \multicolumn{2}{|c}{TELE1}&
    \multicolumn{2}{|c}{TELE2}\\
    \cline{2-13}
    & Acc. & MAE & Acc. & MAE & Acc. & MAE & Acc. & MAE & Acc. & MAE & Acc. & MAE \\
    \midrule
    Hawkes & 0.33 & 1.72 & 0.37 & 1.58 & 0.36 & 2.73 & 0.36 & 1.99 & 0.35 & 2.33 & 0.37 & 2.10\\
    LSTM & 0.37 & - & 0.34 & - & 0.41 & - & 0.39 & - & 0.41 & - & 0.44 & -\\
    SAHP & 0.40 & 1.32 & 0.38 & \underline{1.01} & \underline{0.46} & \underline{2.09} & 0.45 & \underline{1.89} & 0.42 & \underline{2.14} & 0.43 & \underline{1.97}\\
    CT-LSTM & 0.45 & 1.11 & 0.42 & 1.23 & 0.37 & 2.30 & 0.43 & 1.96 & 0.42 & 2.26 & 0.44 & 2.55 \\
    PCT-LSTM & \textbf{0.47} & \underline{1.07} & \underline{0.44} & 1.16 & 0.38 & 2.23 & \underline{0.46} & 1.91 & \underline{0.46} & 2.29 & \underline{0.45} & 2.59 \\
    LOBDIF & \underline{0.46} & \textbf{0.86} & \textbf{0.44} & \textbf{0.84} & \textbf{0.52} & \textbf{1.98} & \textbf{0.50} & \textbf{1.78} & \textbf{0.47} & \textbf{1.96} & \textbf{0.50} & \textbf{1.92}\\
    \bottomrule
    \bottomrule
    p-value & $1.52\text{e-}3$ & $4.27\text{e-}3$ & $3.05\text{e-}3$ & $9.16\text{e-}4$ & $6.10\text{e-}4$ & $3.05\text{e-}3$ & $7.63\text{e-}4$ & $4.82\text{e-}3$ & $1.68\text{e-}3$ & $3.82\text{e-}3$ & $7.63\text{e-}3$ & $1.53\text{e-}3$ \\
    \bottomrule
  \end{tabular}
  }
\end{table*}

\begin{figure*}[t] \centering
\subfigure[MSFT1]{\label{fig:MSFT.time}
\includegraphics[width=0.3\textwidth]{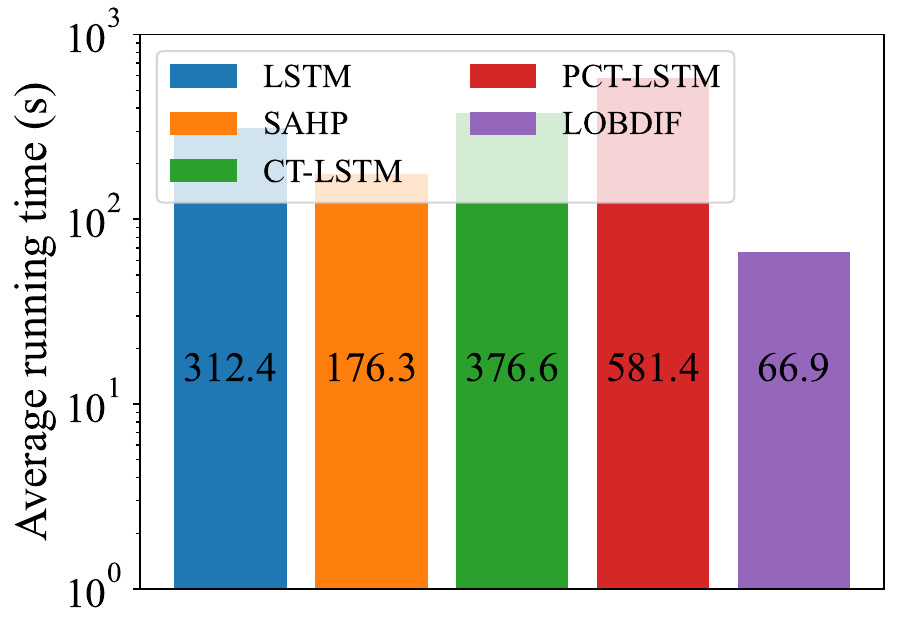}}
\subfigure[PINGAN1]{\label{fig:PINGAN.time}
\includegraphics[width=0.3\textwidth]{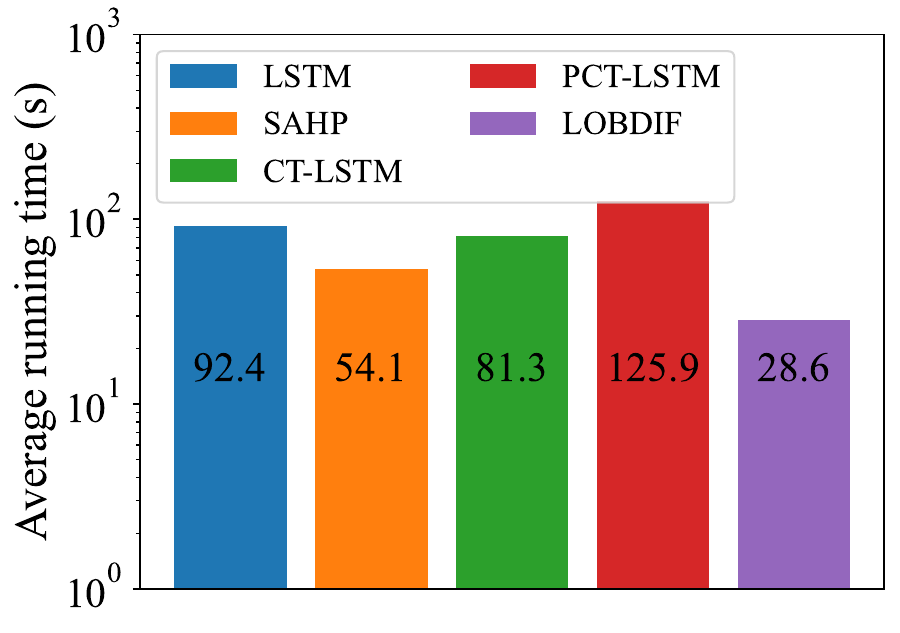}}
\subfigure[TELE1]{\label{fig:TELECOM.time}
\includegraphics[width=0.3\textwidth]{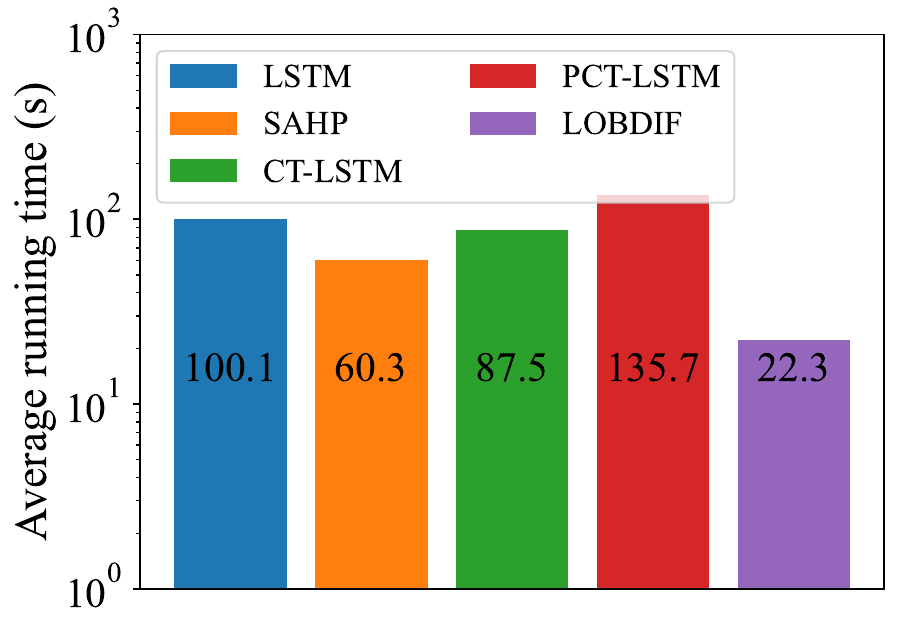}}
\caption{Average running time comparison for testing dataset execution.} \label{fig:time_consume}
\end{figure*}

\subsubsection{Training Setup and Implementation Details}

All experiments are conducted on a machine equipped with an Intel
Xeon Silver 4214R CPU, 256GB RAM and an NVIDIA GeForce RTX 3090
(32GB memory). To ensure reliable results and minimize randomness,
we run all models five times and average their performance. Our
method is compared against several baselines, and their respective
results are reproduced using open-source code with optimal settings
as described in their respective papers. The proposed LOBDIF is
implemented using PyTorch and optimized with the Adam optimizer.
During the training phase, we conduct 200 training epochs with a
learning rate of $2.0 \times 10^{-3}$. The length of the input event
history is set to $L = 50$, and both the time and event encoding
dimensions are set to 64. More details can be found in our
code\footnote{https://github.com/zhengzetao/LOBDIF}.

For the parameter settings of the comparison models, all linear
layers in the LSTM units consist of two layers with 16 units and
Tanh activation, whereas the linear layers used to compute the decay
coefficient utilize Softplus activation. The attention mechanism in
SAHP uses 4 heads. The embedding layers for the event type and state
indicator also comprise two layers with 16 units and Tanh
activation. The decoding layer for intensity rates consists of a
single layer with 16 units and Softplus activation to ensure
positive outputs. All models are optimized using RMSprop with a
learning rate of $2 \times 10^{-3}$ and trained for 200 iterations.

\subsubsection{Evaluation Metrics}

We evaluate performance using the same metrics as in
\cite{DBLP:conf/kdd/ShiC22}: (1) Next event time prediction accuracy
(denoted as \emph{MAE}), calculated as the absolute difference
between the actual and predicted time after applying a common
logarithm, which normalizes the wide range of time spans from
microseconds to seconds; and (2) Next event type prediction accuracy
(denoted as \emph{Acc.}), expressed as a percentage, assessed both
when the next event time is known and unknown.

\subsection{Comparison Results}
\label{sec:exp.compare}

The performance comparison is displayed in Table~\ref{tab:bas_comp},
where the model with the best performance is denoted in bold and
strongest baselines are highlighted with an underline. We observe
that neural network-based models outperform the pure stochastic
Hawkes model across nearly all evaluation metrics. This is because
the stochastic Hawkes model relies on only a few parameters, whereas
neural networks can learn these parameters more effectively,
granting the model greater capacity to capture the dynamics of event
occurrences. The LSTM model, unlike stochastic point process-based
models, treats the event stream as a time series and does not depend
on intensity rates. As a result, it cannot predict event time, and
only event type prediction results are reported here. Notably, LSTM
outperforms the pure stochastic Hawkes model in terms of accuracy,
indicating its advantages in event type prediction. The performance
improvements achieved by SAHP demonstrate that the self-attention
mechanism is more effective for handling long sequences. CT-LSTM and
PCT-LSTM both exhibit strong overall performance compared with all
other alternatives, with PCT-LSTM performing slightly better,
highlighting the advantage of its parallel continuous-time LSTM in
handling event types. Our proposed model, LOBDIF, achieves the best
overall performance compared with all alternatives, though it is
slightly outperformed by PCT-LSTM on MSFT1, still yielding excellent
results. This demonstrates that diffusion-based modeling is
effective in capturing the complex relationships within the order
stream. By decomposing the intricate time-event distribution into
multiple Gaussian distributions, the diffusion model facilitates
more effective optimization and accurate predictions.

Additionally, we perform the paired Wilcoxon signed-rank test to
assess the statistical significance of our method's performance
relative to the strongest baseline. The p-values presented in
Table~\ref{tab:bas_comp} represent the significance level of the
results. The statistical significance analysis is based on 5 pairs
of experimental results from our proposed model and the strongest
baseline, with each model being run 5 times using the same parameter
settings. A confidence level of 0.05, commonly used in other studies
\cite{DBLP:conf/icde/HeCW20}, is also applied in our work. If the
p-value is less than 0.05, it indicates that the performance of our
model is statistically significant compared with the baseline. As
shown in Table~\ref{tab:bas_comp}, all p-values are below 0.05,
confirming that the experimental results are statistically
significant and that our proposed model outperforms the baselines.

\textbf{Time consumption.} We also compare the average time required
to complete testing on the datasets, as shown in
Figure~\ref{fig:time_consume}. For fairness, we exclude the Hawkes
model from this comparison, as it cannot be executed on a GPU. It
can be observed that our model demonstrates superior efficiency
compared with all other models. The average prediction time per step
for the three datasets is 1.5 milliseconds, 1.9 milliseconds, and
1.4 milliseconds, respectively (calculated by dividing the total
testing time by the number of events). This level of efficiency
makes LOBDIF a practical choice for event stream prediction in limit
order book. We attribute this outstanding performance to the
skip-step sampling mechanism, which significantly enhances the
efficiency of the diffusion model. For other models, SAHP exhibits
runtime performance second only to LOBDIF, but the LSTM-based
models, including CT-LSTM and PCT-LSTM, show inferior efficiency.
These models rely heavily on LSTM, which operates as a sequential
chain structure, requiring the output of the previous step to
compute the next step, leading to inherently slower execution.

\subsection{Ablation Study}
\label{sec:exp.ablation}

In this section, we conduct ablation studies to investigate the
impact of various factors on the model performance. Specifically, we
study (1) the influence of the time-event encoder in the denoising
process, (2) the impact of the dedicate network in denoising
process, and (3) the advantage of skip-step sampling in the
prediction phase. We provide detailed explanations of each ablation
study in the following.

\begin{table}[t]
  \caption{Performance comparison of LOBDIF without time and event encodings. TE denotes the time encoding and EE denotes the event encoding.}
  \label{tab:encorder_comp}
  \center
  \resizebox{0.5\textwidth}{!}{
  \begin{tabular}{c|cc|cc|cc}
    \midrule
    \multirow{2}{*}{Datasets}& \multicolumn{2}{c|}{w/o  TE}& \multicolumn{2}{c|}{w/o EE} & \multicolumn{2}{c}{w/o TE \& EE} \\
    \cline{2-7}
      & Acc. & MAE & Acc. & MAE & Acc. & MAE\\
    \midrule
    MSFT1 & -4.1\% & -3.4\% & -2.3\% & -1.9\% & -5.1\% & -6.4\% \\
    MSFT2 & -3.6\% & -3.4\% & -1.9\% & -2.3\% & -6.4\% & -5.5\% \\
    \midrule
    PINGAN1 & -3.7\% & -2.8\% & -0.9\% & -0.1\% & -2.8\% & -1.9\%  \\
    PINGAN2 & -3.1\% & -3.2\% & -1.3\% & -0.6\% & -3.9\% & -3.3\%  \\
    \midrule
    TELE1 & -2.7\% & -1.9\% & -1.0\% & -3.6\% & -4.0\% & -4.1\%  \\
    TELE2 & -3.2\% & -4.5\% & -2.1\% & -2.6\% & -5.3\% & -4.7\%  \\
    \bottomrule
  \end{tabular}}
\end{table}

\subsubsection{Influence of Time-event Encoder}

We conduct a detailed analysis of the roles of time encoding and
event encoding in the model. The experimental results are presented
in Table~\ref{tab:encorder_comp}. For clarity, we compare the
results of different configurations with the full LOBDIF model
results from Table~\ref{tab:encorder_comp} and display the
differences as percentages. When the time encoding module is
removed, the model uses the raw time values concatenated with event
encodings as input. Conversely, when the event encoding module is
removed, the model uses time encodings along with the one-hot
embeddings of events as input. Overall, both modules are
indispensable for LOBDIF's performance. Specifically, time encoding
has a more significant impact on the model's performance, while the
effect of event encoding is relatively smaller. This highlights the
critical role of time encoding in handling event stream with
irregular time point.

\subsubsection{Impact of Denoising Network}

To effectively capture the interdependence between time and events,
we design a specialized denoising network. To validate its
effectiveness, we replace this carefully designed network with MLP
and GRU networks, and the experimental results are shown in
Table~\ref{tab:ablation_net}. Due to space limitations, we present
results for only three datasets.

As observed in Table~\ref{tab:ablation_net}, our carefully designed
network achieves superior performance in both metrics across the
three datasets. This is primarily because our network leverages time
attention and event attention to effectively capture the
interdependence between temporal and event features. In contrast,
the other two networks perform poorly. While MLP and GRU can extract
either event features or temporal features, their effectiveness is
limited to regularly sampled time series. For complex and
irregularly sampled time sequences, such as event stream in LOB, it
is crucial to specifically capture the intricate interdependence
between events and time.

\begin{table}[t]
  \caption{Performance comparison of LOBDIF with other denoising networks.}
  \label{tab:ablation_net}
  \center
  \resizebox{0.4\textwidth}{!}{
  \begin{tabular}{c|cc|cc}
    \midrule
    \multirow{2}{*}{Datasets}& \multicolumn{2}{c|}{MLP}& \multicolumn{2}{c}{GRU} \\
    \cline{2-5}
      & MAE & Acc. & MAE & Acc.\\
    \midrule
    MSFT1 & -6.8\% & -3.2\% & -3.8\% & -4.5\% \\
    MSFT2 & -5.6\% & -4.0\% & -2.9\% & -4.7\% \\
    \midrule
    PINGAN1 & -7.2\% & -5.5\% & -4.9\% & -6.3\% \\
    PINGAN2 & -6.7\% & -5.1\% & -3.2\% & -4.6\% \\
    \midrule
    TELE1 & -5.0\% & -3.9\% & -2.7\% & -4.3\% \\
    TELE2 & -4.6\% & -3.5\% & -3.0\% & -4.9\%  \\
    \bottomrule
  \end{tabular}}
\end{table}

\begin{figure*}[t] \centering
\subfigure[MSFT1]{\label{fig:MSFT.skip}
\includegraphics[width=0.3\textwidth]{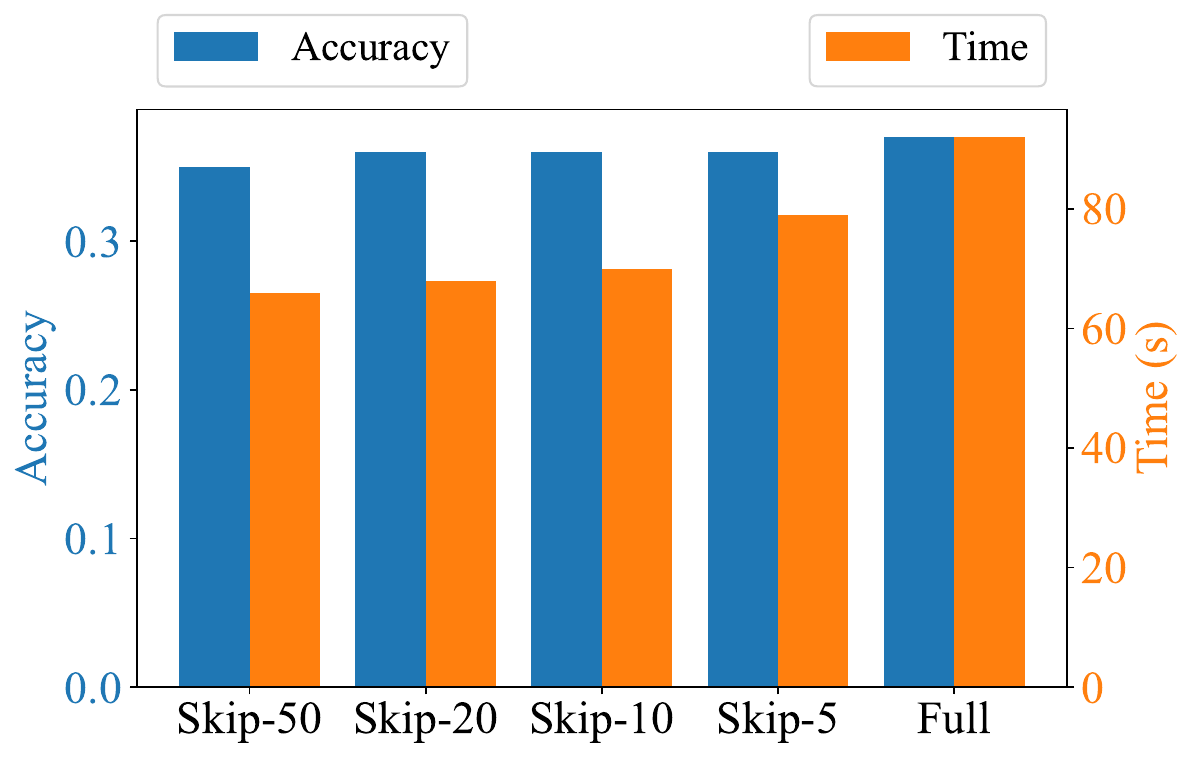}}
\subfigure[PINGAN1]{\label{fig:PINGAN.skip}
\includegraphics[width=0.3\textwidth]{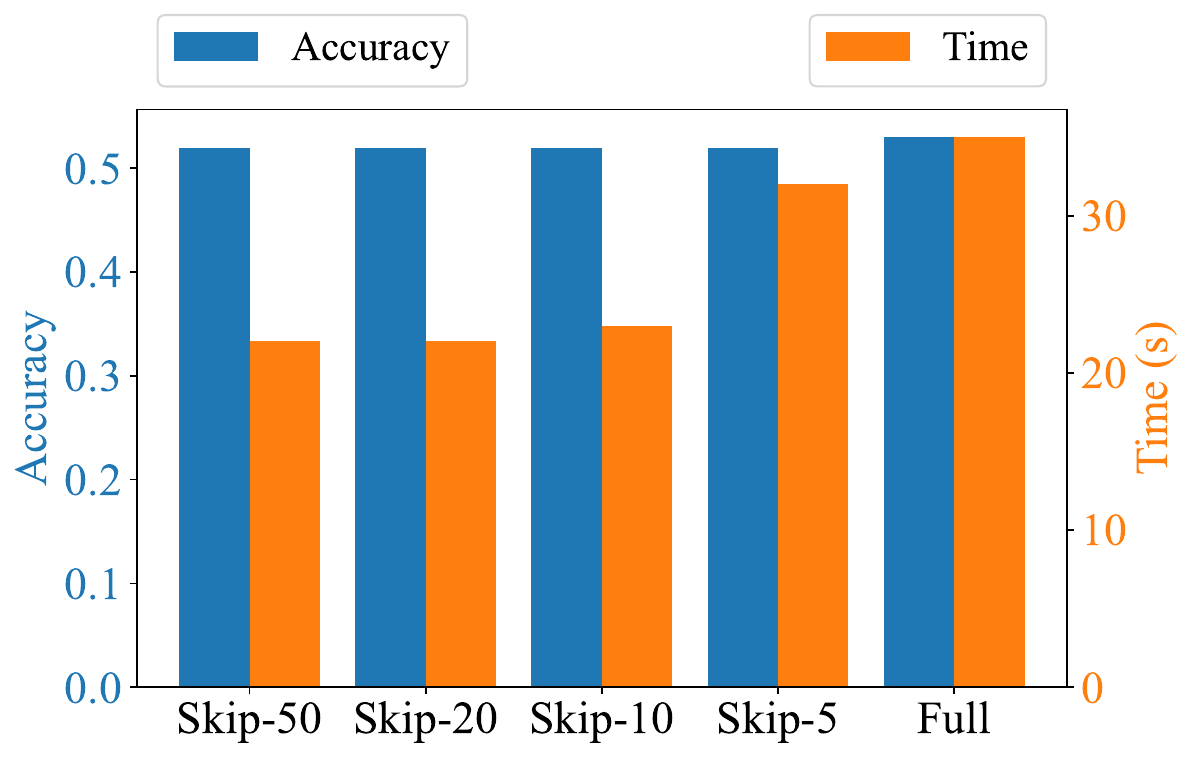}}
\subfigure[TELE1]{\label{fig:TELECOM.skip}
\includegraphics[width=0.3\textwidth]{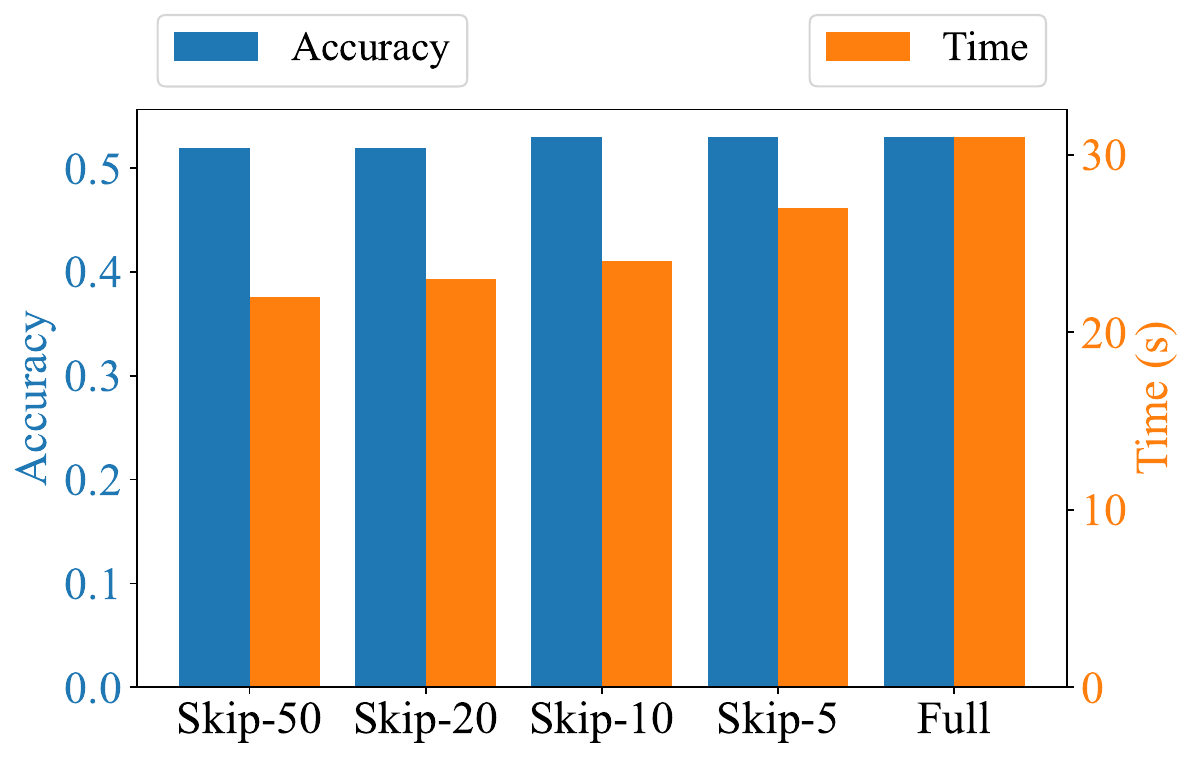}}
\caption{Average running time comparison for testing dataset
execution.} \label{fig:skip_exp}
\end{figure*}

\subsubsection{Advantage of Skip-step Sampling}

We introduce skip-step sampling to accelerate the denoising process,
making it more suitable for LOB event stream prediction scenarios.
The model no longer executes a step-by-step denoising process but
instead performs skip-step sampling to enhance efficiency, enabling
predictions to be completed in fewer steps. In this experiment, we
set the number of skipped steps as the parameter $\tau$. The
parameter $\tau$ determines how many steps are skipped during the
denoising process, thereby influencing the speed of the process. A
larger $\tau$ value means skipping more time steps during denoising,
which significantly accelerates the prediction process.

The value of $\tau$ plays a crucial role in the model's performance.
Under a forward diffusion setup with $K = 100$ steps, we explore the
impact of different $\tau$ values ($\tau = \{5, 10, 20, 50\}$) with
regard to the model's efficiency and effectiveness. The results,
presented in Figure~\ref{fig:skip_exp}, show that as the number of
denoising steps decreases, the model's prediction speed improves
significantly. However, this reduction in denoising steps has almost
no impact on prediction accuracy, suggesting that the model can
maintain performance even with smaller numbers of steps in the
denoising process. This indicates that skip-step sampling
effectively accelerates the inference process without sacrificing
the quality of the predictions. Nevertheless, the speed-up achieved
by skip-step sampling is critical for real-time prediction tasks,
particularly for large-scale LOB data stream predictions.

\begin{figure*}[t] \centering
\subfigure[Diffusion steps analysis]{\label{fig:hyper.skip}
\includegraphics[width=0.3\textwidth]{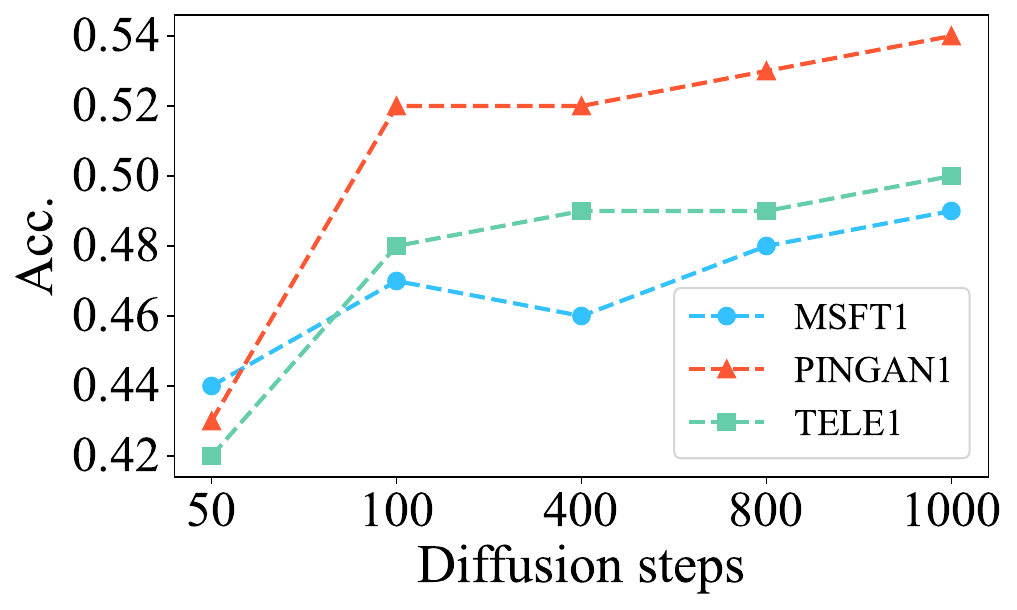}}
\subfigure[Training epochs analysis]{\label{fig:hyper.epoch}
\includegraphics[width=0.3\textwidth]{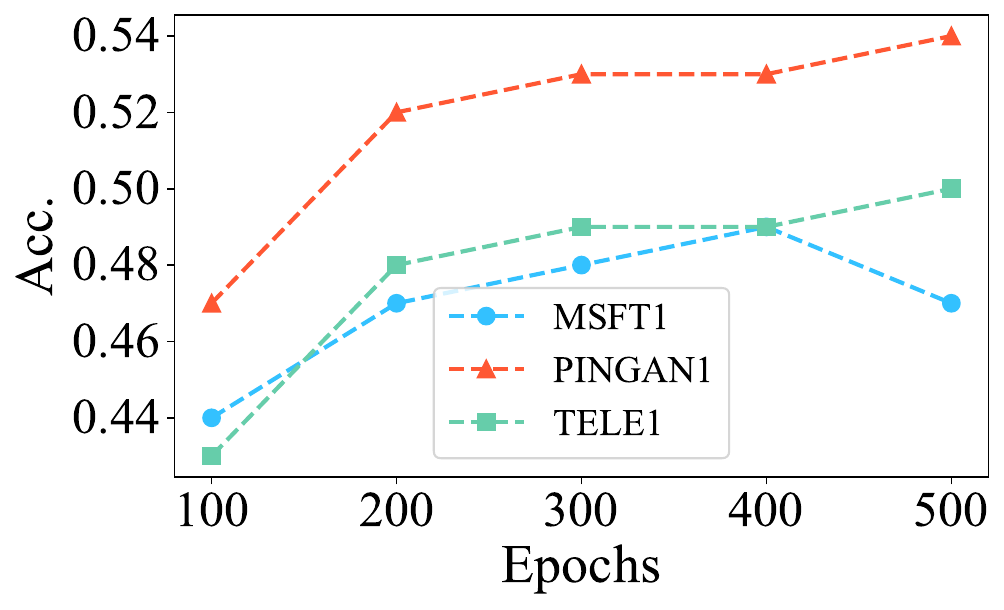}}
\subfigure[Encoding dimensions analysis]{\label{fig:hyper.dimen}
\includegraphics[width=0.3\textwidth]{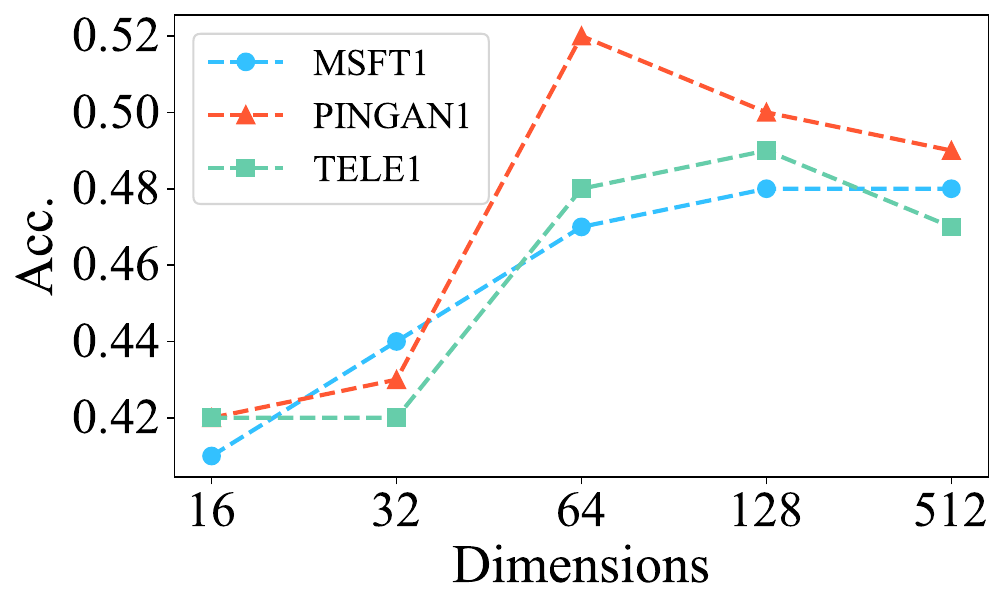}}
\caption{Analysis of the proposed model's performance with different
hyperparameters.} \label{fig:hyper_exp}
\end{figure*}

\begin{figure*}[t] \centering
\subfigure[MSFT1]{\label{fig:msft}
\includegraphics[width=0.95\textwidth]{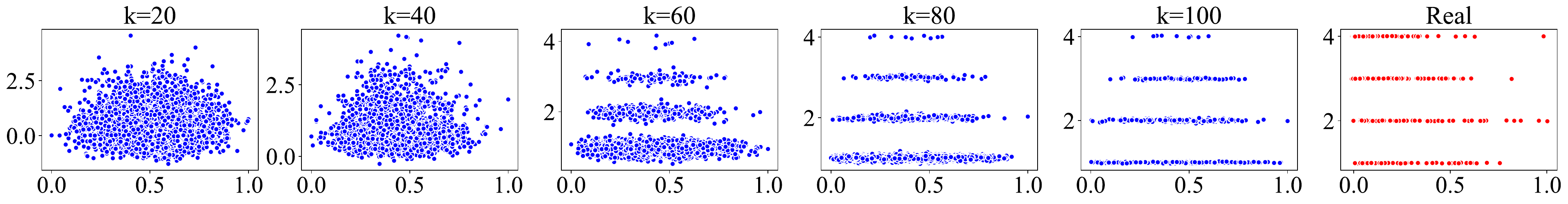}}
\subfigure[PINGAN1]{\label{fig:pingan}
\includegraphics[width=0.95\textwidth]{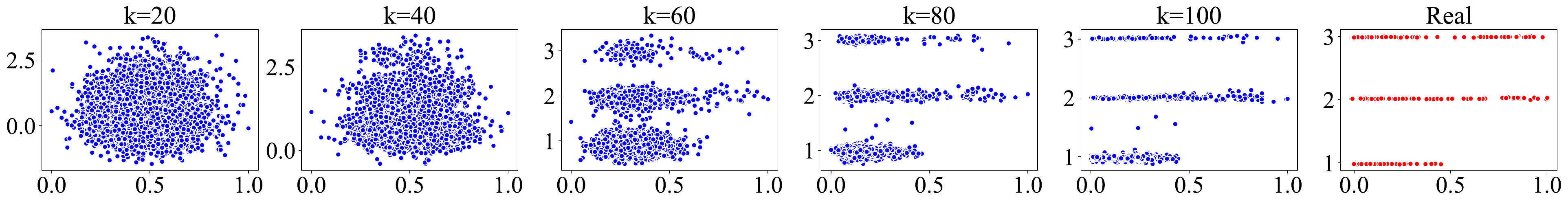}}
\subfigure[TELE1]{\label{fig:telecom}
\includegraphics[width=.95\textwidth]{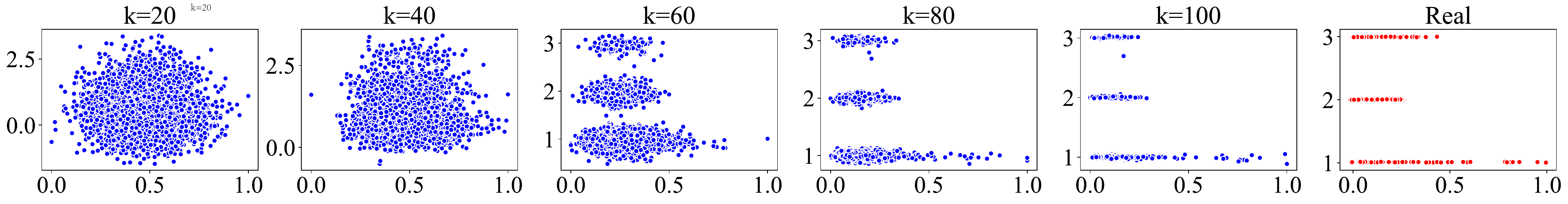}}
\caption{Visualization of the time-event distribution. The first
five columns (in blue) illustrate different stages of the denoising
process, while the final column (in red) represents the true
time-event distribution. Starting from Gaussian noise, our model
gradually approximates the ground-truth distribution}
\label{fig:distribution}
\end{figure*}

\subsection{Hyperparameter Analysis}
\label{sec:exp.hyper}

Many hyperparameter settings significantly impact the performance of
our model. Key factors such as the number of diffusion steps $K$,
the number of training epochs, and the dimensions of the time-event
encoding play critical roles in determining the model's
effectiveness. We analyze the model's performance under different
parameter configurations to discuss optimal parameter selection,
with the results shown in Figure~\ref{fig:hyper_exp}.

\subsubsection{Impact of Diffusion Step $K$}

The number of diffusion steps, $K$, plays a critical role in the
model's performance. A larger $K$ allows for a more detailed and
granular diffusion process, potentially leading to better
predictions. To investigate its impact, we experiment with different
values of $K = \{50, 100, 400, 800, 1000\}$, analyzing the model's
performance and the training time consumption. The results are
presented in Figure~\ref{fig:hyper.skip}. From the figure, we
observe that as $K$ increases, the model's predictive accuracy
improves. This improvement occurs because a larger $K$ enables the
model to better approximate the true data distribution, particularly
for complex time-event relationships in the event stream. However,
this comes at a cost: the training time per epoch increases
significantly with $K$. This is because each diffusion step involves
both the forward pass and the computation of gradients, and a larger
$K$ increases the number of computational steps required during
training.

While increasing $K$ can improve results, there is a diminishing
return in accuracy beyond a certain point. For example, the
performance gain from $K = 400$ to $K = 800$ is much smaller
compared with the gain from $K = 50$ to $K = 100$. This indicates
that excessively large $K$ values may lead to diminishing benefits
while disproportionately increasing computational cost. Therefore,
selecting an appropriate $K$ value is crucial for balancing model
accuracy and efficiency. For practical applications such as
real-time predictions in limit order books, a moderate $K$ may
provide the best trade-off between computational feasibility and
prediction quality.

\subsubsection{Impact of Training Epochs}

We set the model's training epochs to values in the range $\{100,
200, 300, 400, 500\}$, with the results illustrated in
Figure~\ref{fig:hyper.epoch}. Increasing the number of training
epochs generally improves model performance, as the model is able to
better learn the underlying patterns in the data. However, the
performance gain becomes less significant beyond a certain point.
For example, while the improvement from 100 to 300 epochs is
notable, the difference between 400 and 500 epochs is minimal. This
suggests that additional training epochs yield diminishing returns
once the model approaches its optimal learning capacity.

Moreover, increasing the number of epochs inevitably leads to higher
training time consumption. Considering that the forward diffusion
process itself is computationally intensive, excessively large epoch
values are not practical, especially for time-sensitive
applications. Instead, we prioritize a balanced approach, selecting
$\text{epoch} = 200$ as an optimal trade-off. This setting achieves
good performance without incurring excessive computational cost.

\subsubsection{Impact of Encoding Dimension}

We experiment with different encoding dimensions for the model,
setting the values to $\{16, 32, 64, 128, 512\}$. The results, shown
in Figure~\ref{fig:hyper.dimen}, demonstrate how the encoding
dimension affects the model's performance. Increasing the encoding
dimension generally enhances the vector's representation ability,
allowing the model to capture more complex patterns in the data.
However, as observed in the figure, the performance improvement
becomes less significant as the dimension increases beyond 64.
Notably, the model's performance with dimensions of 64 is comparable
to that with 256 dimensions. This performance can be attributed to
the nature of time and event type data, which do not carry as much
complexity or richness as data in other domains, such as natural
language or images. Excessively high dimensions may lead to
overparameterization, increasing computational costs without
meaningful performance gains. For instance, larger dimensions
require more memory and longer training time, which are impractical
for applications requiring real-time processing. Considering these
factors, we select an encoding dimension of 64 as a balanced choice.
It provides sufficient expressive power to capture the key patterns
in the time and event type data while maintaining computational
efficiency.

\subsection{Case Study on Denoising Progress}
\label{sec:exp.case}

To gain a deeper understanding of the denoising process, we
visualize the time-event distribution during the reverse denoising
iterations in Figure~\ref{fig:distribution}. For better
visualization, we normalize the intervals between consecutive
timestamps to the range (0, 1). Additionally, only 5000 event points
are selected for a clearer representation, with 100 diffusion steps
(as a larger number of points would make the visualization too
dense).

As shown in Figure~\ref{fig:distribution}, at the start of the
denoising process, the time-event distribution appears as Gaussian
noise. With each progressive denoising iteration, the event type
distribution gradually converges towards different event types,
while the event time spreads across different time distributions. By
the final step, the time-event distribution closely matches the
ground-truth distribution, indicating that our LOBDIF model
effectively learns the complex time-event relationships.

Moreover, the quality of the denoising improves throughout the
iterations, as both the time and event type distributions gradually
align with the true distribution. This suggests that the
interdependence between the time and event domains is effectively
captured in the process of denoising, which explains the significant
improvement in performance during this period.

\section{Conclusion}
\label{sec:conclu}

In this paper, we explore the use of diffusion models for limit
order book event stream prediction for the first time. Specifically,
we propose a novel model named LOBDIF, which breaks away from the
traditional stochastic probability-based prediction methods by
leveraging a diffusion model for event stream prediction. The
step-by-step mechanism of diffusion models facilitates the
decomposition of the complex target time-event distribution into
simpler Gaussian distributions, enabling a more effective capture of
the interdependence between time and event types. To effectively and
efficiently model the relationship between time and event types, we
introduce two key components: a denoising network and a skip-step
sampling strategy. The denoising network excels at capturing the
intricate patterns between time and event types, while the skip-step
sampling strategy accelerates the denoising process during
prediction, significantly improving efficiency. Extensive
experiments demonstrate that our model achieves superior
effectiveness and efficiency compared to state-of-the-art methods,
validating its potential for limit order book event stream
prediction.

\bibliographystyle{IEEEtran}
\bibliography{IEEEtran}

\end{document}